\documentclass[prd,
               twocolumn,
               eqsecnum,
               showpacs,
               letterpaper,
               superscriptaddress,
               altaffilletter]{revtex4-1}
 \usepackage{graphicx}
\usepackage{amsmath,amssymb}
\usepackage[latin1]{inputenc}
\usepackage{graphicx}
\usepackage[english]{babel}
\usepackage{amsbsy}
\usepackage{textcomp}
\usepackage{psfrag}
\usepackage{multirow}
\usepackage{color} 
\newcommand{\ud}{\mathrm{d}}    
\def\urltilda{\kern -.15em\lower .7ex\hbox{\~{}}\kern .04em}

\begin{document}

\title{Application of asymptotic expansions for maximum likelihood estimators errors to gravitational waves from binary mergers: the single interferometer case. }
\author{M. Zanolin}
\affiliation{Embry-Riddle Aeronautical University, 3700 Willow Creek Road, Prescott, AZ, 86301,USA}
\author{S. Vitale}
\affiliation{Embry-Riddle Aeronautical University, 3700 Willow Creek Road, Prescott, AZ, 86301,USA}
\affiliation{Universit\'e Pierre-et-Marie-Curie - 4, Place Jussieu, 75005 Paris, France}
\author{ N. Makris}
\affiliation{Massachusetts Institute of Technology, 77 Mass Ave, Cambridge, MA, 02139, USA}

 \begin{abstract}
In this paper we apply to gravitational waves (GW) from the merger phase of 
binary systems a recently derived frequentist methodology to calculate analytically
the error for a maximum likelihood estimate (MLE) of physical parameters.
We use expansions of the covariance and the bias of a MLE estimate 
in terms of inverse powers of SNRs where the square root of the first order
in the covariance expansion is the Cramer Rao Lower Bound (CRLB). 
We evaluate the expansions, for the first time, for GW signals in noises of GW interferometers.
The examples are limited to a single, optimally oriented, interferometer. 
We also compare the error estimates using the first two orders of the expansions
 with existing numerical Monte Carlo simulations.
The  first two orders of the covariance allows to get error predictions closer to what is 
observed in numerical simulations than the CRLB. 
The methodology also predicts a necessary SNR to approximate the error
with the CRLB and provides new insight on the
relationship between waveform properties, SNR, dimension of the parameter space
 and estimation errors.
For example the timing match filtering can achieve the CRLB only if the SNR
is larger than the Kurtosis of the gravitational wave spectrum and the necessary SNR 
is much larger if other physical parameters are also unknown.   
\end{abstract}

\maketitle


\section{Introduction}
The ground-based gravitational waves detectors LIGO, Virgo, and GEO 600
(\cite{Abbott2009}, \cite{Grote2008}, \cite{Acernese2006})
are rapidly improving in sensitivity.
By 2015, advanced versions of these detectors should be taking
data with a design sensitivity approximately 10 times greater than the previous generation, and the probed volume will grow
by a factor of about a thousand. Such improvements in detector sensitivity mean that the first gravitational-wave
signature of a compact-binary coalescence (CBC)
could be detected in the next few years (see for example \cite{Kalogera2007}).
Among the expected signals, a special role is covered by inspiralling compact binaries. This follows
from the ability to model the phase and amplitude of the signals quite accurately and consequently to maximize the signal-to-noise ratio (SNR) by using matched filtering techniques.
Matched filters also provide a maxim likelihood estimation of the waveform parameters
such as component masses or time of coalescence.
The choice of the MLEs as reference estimators is also motivated by the fact that if an unbiased estimator that attain the CRLB exists, is the MLE\cite{willsky}.\\
\\
The existing GW frequentist literature (\cite{Arun1}..\cite{Cokelaer}) evaluates the MLE accuracy in two ways: (a) analytically by calculating the
so called Fisher information matrix (FIM) or equivalently the Cramer Rao Lower Bound (CRLB) which is the square root of the 
diagonal elements of the inverse FIM, and (b) numerically by performing Monte Carlo simulations. \\
The FIM  was derived analytically in (\cite{Cutler},\cite{Finn1993},\cite{Jaranowski1994}) using Newtonian waveforms, extended to second-order post-Newtonian (\cite{Krolak},\cite{Poisson}) and recently revisited up to 3.5PN (\cite{Arun1},\cite{ArunError}).
In (\cite{Arun1}),(\cite{ArunError}) the authors calculate the CRLB for the three {\it standard} binary systems (NNS, NBH, BBH), and show how the errors change when the successive different PN orders are taken into account. They consider initial LIGO, advanced LIGO and Virgo noises. They also considers PN corrections to the {\it amplitude}.\\
Monte Carlo simulations were performed, for example in (\cite{Bala1996}, \cite{Bala1998}), for the lowest PN orders, where it is also suggested that the inclusion of the higher PN orders would be computationally expensive.
More recent Monte Carlo simulations  with 3.5 PN waveforms
are described in (\cite{Cokelaer}).\\
We did not try to compare the uncertainties derived here to other existing papers (especially from the '90)
since different parameter sets, noise spectra (even for the same antenna) and PN terms were used.
For example in  (\cite{Vecchio}) a comparison between the CRLB and other bounds is done for a waveform
at the 0th PN order. This work also uses different conventions
on the waveform spectrum than more recent literature.
In (\cite{Cutler}) phasing is extended to the 1.5PN order. The spin parameters are taken in account but
the noise spectrum for LIGO is different than the currently used design noise curves.
In (\cite{Poisson}), (\cite{Krolak}), (\cite{Bala1996}) the 2PN wave was used.
In the work (\cite{Poisson}) interesting observations are made about the fluctuation of the parameters variance with the PN orders, analyzing both the case of spin and spinless systems. The fluctuations of the variance in the spinless case is also stressed in (\cite{Arun1}).\\
\\
The CRLB is a convenient tool to approximate the accuracies in large SNRs and to obtain error bounds for unbiased estimators. Unfortunately, for low SNRs (below 20) where the first detections might emerge, the CRLB can grossly underestimate the errors \cite{Vecchio},\cite{Vallisneri2008},\cite{cutlerflanagan1994},\cite{finn1992},\cite{Bala1996}.
The reason is that with non linear data models and (or) with non Gaussian noises the CRLB depends only on the curvature of the likelihood function around the true value of the parameter.\\
\\
In this paper we apply a recently derived analytical tool to better predict an MLE accuracy and
to establish necessary conditions on the signal-to-noise ratio (SNR) for the MLE error to attain the CRLB.
Explicitly, within the frequentist framework, for arbitrary probability
distribution of the noise, expansions of the bias and the covariance of a MLE
in inverse powers of the SNR are discussed. The first order of the expansion of the
variance is the inverse of the FIM. By requiring that the second order covariance 
is smaller, or much smaller, than the first order this approach predicts a necessary SNR 
to approximate the error with the CRLB. The higher order of the expansions are determinant 
in the low SNR regime where the inverse FIM  underestimates the error. 
We compared our the errors computed using the first two orders of the expansions 
to the Monte Carlo simulations in
(\cite{Cokelaer}). We observed the first two orders of the expansions get error predictions closer 
than the CRLB to what is observed in the numerical simulations.
In (\cite{Cokelaer}) the simulations are related to the FIM to estabish ranges of SNR where 
the CRLB describe the error. Our expansions predict the same SNR range of validity for the CRLB.
\\
The expansions are 
sensitive to the side lobes of the likelihood function because they make use of higher order 
derivatives than the second one (which is only sensitive to the curvature of the main lobe).
The methodology also provides new insight on the
relationship between waveform properties, SNR, dimension of the parameter space
 and estimation errors. For example the timing match filtering accuracy achieves the CRLB only if the SNR
is larger than the Kurtosis of the gravitational wave spectrum and the necessary SNR
is much larger if the other physical parameters are unknown.
More specifically the MLE of the arrival time for NS-NS binary signals 
might require an SNR equal to 2 with the time as the only parameter or 15 when all the other parameters are unknown.
These results are important to understand the domain of validity of recent papers like 
\cite{fairhust}  that defines $90\%$ confidence regions in direction reconstruction
with time triangulation. The regions discussed in \cite{fairhust} for SNR smaller than 10 
are based on timing MLEs, with the arrival time being the only unknown parameter, and the 
time uncertainties quantified by the CRLB.\\
We also note that
\cite{Vallisneri2008}, using a formalism introduced in \cite{cutlerflanagan1994},\cite{finn1992},
describes a consistency criterion, different from the condition derived in this paper,
 for the validity of the CRLB that, if applied to a 4pp compact binary signal
computed with a 2PN expansion and $m_1=m_2=10 M_{\odot}$, requires an SNR of at least 10.
At the time of the writing of this paper we established with M.Vallisneri that 
the equation (\ref{VarMatrixSimplified}) of this paper becomes, in the one parameter case and colored Gaussian noise,
equivalent to equation (60) in (\cite{Vallisneri2008}) or (A35) in (\cite{cutlerflanagan1994}).
A comparison for the Gaussian noise and multi parameter case is object of current work, while a comparison for 
arbitrary noise is not possible because (\cite{Vallisneri2008}) and (\cite{cutlerflanagan1994}) 
use from the beginning of their derivations Gaussian noises. The explicit calculations shown here 
for different GWs are also not performed in (\cite{Vallisneri2008}) and (\cite{cutlerflanagan1994}).\\
\\
In section \ref{Expansions} we present the explicit expressions of the expansions of the bias and the covariance matrix
for arbitrary noise and size of the parameter space. In section \ref{Match} we explain how the expansion can be evaluated for signals in additive colored Gaussian noise.
In section \ref{WaveformSection} we describe the post-Newtonian inspiral waveform used for the examples, describe the parameter space and the initial and advanced LIGO noises.
In section \ref{OneDim} we study the one-dimensional parameter space results when only one parameter at a time is considered unknown. 
In section \ref{fullparspace} We present the results for full parameter space with the initial and advanced LIGO noises. We also compare our results with published results from Monte Carlo simulations.
In section \ref{conclusions} we present some conclusions and in the appendix we describe the derivation of the 
expansions as well as the relationship of this method with the statistics literature.
\section{Expansions for the bias and covariance matrix of a frequentist MLE in arbitrary noise}\label{Expansions}
In this section we present the first two orders of the expansions in inverse powers of the SNR
for the bias and the covariance matrix. The details of the derivation are provided in appendix A.
Given an experimental 
data vector ${\underline{x}}=\{x_{1},..,x_{N} \}$,
where $N$ is the dimension, we assume that the data are described by a probability density 
$P(\underline{{x}},\underline{\vartheta})$ that depends 
on a D-dimensional parameter vector $ \underline{{\vartheta}}=\{
{{\vartheta}}_1,..,{{\vartheta}}_D \}  $.
According to (\cite{Lawley}), we suppose that the MLE $ \underline{\widehat{\vartheta}}=\{
{\widehat{\vartheta}}_1,..,{\widehat{\vartheta}}_D \}  $
of  $\underline{\vartheta}$ is given by a stationary point of the  
likelihood function $l(\underline{x},\underline{\vartheta}) 
=ln(P(\underline{x},\underline{\vartheta}))$ with respect to the components of  
$\underline{\vartheta}$ 
\begin{eqnarray}
  l_r(\underline{x},\underline{\widehat{\vartheta}}) =\label{stazionarieta0} \frac{\partial l(\underline{x},\underline{\vartheta})}{\partial\vartheta_r }{{\mid}_{\underline{\vartheta}=\underline{\widehat{\vartheta}}}}=0 \,\,\,\,\,\,\, r=1,..,D.
\end{eqnarray}
If we introduce the notations 
 \begin{equation*}
l_{a_1a_2..a_s} =l_{a_1a_2..a_s}(\underline{x},
\underline{\vartheta})= \frac{\partial^s l(\underline{x},
\underline{\vartheta})}{\partial\vartheta_{a_1}\partial\vartheta_{a_2}..\partial\vartheta_{a_s} }
\end{equation*}
\begin{equation*}
 \upsilon_{a_1 a_2..a_s ,\,\,..\,\,, b_1 b_2..b_s}=E[l_{a_1 a_2.. a_s}\,\,..\,\,l_{b_1 b_2..b_s}]  
\end{equation*}


where  $-\upsilon_{ab}$ is the Fisher information matrix
 $i_{ab} =-\upsilon_{ab}= -E[l_{ab}]=E[l_{a}l_{b}]$ ($E[.]$ is the expected value),
the first two orders of the bias for the MLE of the $r$ component of the 
parameter vector $\underline{\vartheta}$ are given by
\begin{eqnarray}\label{BiasOne}
b_1(\widehat{\vartheta }^m)&=&\frac{1}{2}i^{ma}i^{bc}(\upsilon_{abc}+2\upsilon_{c,ab})
\end{eqnarray}
\begin{widetext}
 \begin{eqnarray}\label{BiasTwo}
b_2(\widehat{\vartheta}^m) &=& -\frac{i^{ma}i^{bc}}{2}[v_{abc} + 2v_{ab,c}] + \frac{i^{ma}i^{bd}i^{ce}}{8}[v_{abcde} + 4v_{ac,bde} + 8v_{de,abc} + 4v_{abce,d} + 4v_{abc,d,e}\nonumber\\
&+& 8v_{ab,cd,e}]+ \frac{i^{ma}i^{bc}i^{df}i^{eg}}{4}\bigg[(2v_{afed}v_{gb,c} + 2v_{bedf}v_{ac,g} + 4v_{abed}v_{gf,c}) + (v_{afed}v_{gcb} +\nonumber\\
&+& 2v_{abed}v_{gcf} + 2v_{dbeg}v_{acf}) + (2v_{aed}v_{gb,fc} + 4v_{acf}v_{dg,eb} + 4v_{bed}v_{ac,gf} + 2v_{fcb}v_{ag,ed}) +\nonumber\\
&+& (4v_{afe,g}v_{db,c} + 4v_{afe,c}v_{db,g} + 4v_{dbe,g}v_{af,c}) + (2v_{abe,g}v_{cdf} + 4v_{dbe,g}v_{acf} + 4v_{abe,f}v_{cfg} +\nonumber\\ &+& 2v_{dge,b}v_{acf}) + (4v_{ag,fc}v_{ed,b} + 4v_{ed,fc}v_{ag,b} + 4v_{ag,ed}v_{fc,b}) \nonumber \\
&+& \left.(4v_{acg}v_{ef,b,d} + 2v_{cde}v_{ab,f,g}) + \frac{2}{3}v_{abde}v_{c,f,g}\right] \nonumber \\
&+& \frac{i^{ma}i^{bc}i^{de}i^{fg}i^{ti}}{8}[v_{adf}(v_{ebc}v_{gti} + 2v_{etc}v_{gbi} + 4v_{gbe}v_{tci} + 8v_{gbt}v_{eci} + 2v_{ebc}v_{gt,i} \nonumber \\
&+& 4v_{etc}v_{gb,i} + 2v_{gti}v_{eb,c} + 4v_{gtc}v_{eb,i} + 8v_{gbt}v_{ce,i} + 8v_{gbt}v_{ci,e} + 8v_{gbe}v_{ct,i} + 8v_{cte}v_{gb,i} \nonumber \\
&+& 4v_{cti}v_{gb,e} + 4v_{gt,i}v_{eb,c} + 4v_{eb,i}v_{gt,c} + 8v_{gt,b}v_{ic,e} + 8v_{gt,e}v_{ic,b} + 4v_{bet}v_{g,c,i}) \nonumber \\
&+& v_{dci}(8v_{bgt}v_{ae,f} + 4v_{bgf}v_{ae,t} + 8v_{ae,t}v_{bg,f} + 8v_{ae,f}v_{bg,t} + 8v_{af,b}v_{ge,t})]
\end{eqnarray} 
\end{widetext}

were we assumed the Einstein convention to sum over repeated indices.
For the covariance matrix the first order is the inverse of the Fisher information matrix while the second order is given in by (for simplicity we provide the diagonal terms):

\begin{widetext}
\begin{eqnarray}\label{VarMatrix}
C_2(\vartheta^j)\!&=&\!-i^{jj}
\!+\!\label{cocv2}i^{jm}i^{jn}i^{pq}(2\upsilon_{nq,m,p}\!+\!\upsilon_{nmpq}\!+\!3\upsilon_{nq,pm}\!+\!2\upsilon_{nmp,q}\!+\!
\upsilon_{mpq,n})\!+\nonumber\\
&+&i^{jm}i^{jn}i^{pz}i^{qt}\bigg[(\upsilon_{npm}\!+\upsilon_{n,mp})(\upsilon_{qzt}+2\upsilon_{t,zq})+\upsilon_{npq}\left(\frac{5}{2}\upsilon_{mzt}+2\upsilon_{m,tz}+\upsilon_{m,t,z}\right)\nonumber\\
&+&\upsilon_{nq,z}(6\upsilon_{mpt}+2\upsilon_{pt,m}+\upsilon_{mp,t})\bigg]
\end{eqnarray}
\end{widetext}

\section{Expansions for signals in additive colored Gaussian noise}\label{Match}

For this analysis we assume that the output of an optimally oriented GW interferometer has the form:

\begin{equation}\label{output}
 x(t)= h(t,\underline{\theta})+w(t)
\end{equation}
where $h(t,\underline{\theta})$ is the signal, which depends on the parameters vector $\underline{\theta}$, and $w(t)$ a stationary Gaussian noise with zero mean. 
The probability distribution can be written:
\begin{widetext}
\begin{equation}
 p(x) \propto \exp\left\{-\frac{1}{2}\,\int{[x(t)-h(t,\underline{\theta})]\,\Omega(t-t_1)\,[x(t_1)-h(t_1,\underline{\theta})]\,\ud t \ud t_1}\right\}
\end{equation}
\end{widetext}
The first and second derivative of the log-likelihood give:
\begin{widetext}
\begin{eqnarray}
 l_a &\equiv& \frac{\partial \log {p(x)	}}{\partial \theta_a} =  \int{h_a(t,\underline{\theta})\,\Omega(t-t_1)\,[x(t_1)-h(t_1,\underline{\theta})]\,\ud t \ud t_1}\\
l_{ab} &\equiv& \frac{\partial l_a}{\partial \theta_b} =  \int{\left[h_{ab}(t,\underline{\theta})\,\Omega(t-t_1)\,[x(t_1)-h(t_1,\underline{\theta})] - h_a(t,\underline{\theta})\,\Omega(t-t_1)\,h_b(t_1,\underline{\theta}) \right]\ud t \ud t_1}
\end{eqnarray}
\begin{eqnarray}\label{FisherFreqOld}
 i_{a\,b} = E[l_a\, l_b] = -E[l_{ab}]&=& \int{h_a(t,\underline{\theta})\,\Omega(t-t_1)\,h_b(t_1,\underline{\theta})\,\ud t \ud t_1} =\\
&=& \int{\ud f \ud f' \,h_a(f) \,h_b(f') \,\Omega(-f,-f')}
\end{eqnarray}
\end{widetext}
It's easy to verify that 
\begin{eqnarray}
\Omega(f,f')= \frac{\delta(f+f')}{S_h(f)}\label{sign}
\end{eqnarray}
where we have introduced, $S_h(f)$,the {\it one sided power spectral density} defined as the Fourier transform of the noise auto-correlation:
\begin{eqnarray}
 R(t)&=& E[n(t+\tau) n(\tau)].\\
S_h(f)&\equiv& \int{\ud t\, e^{-2\pi i f t} R(t)}.\label{OneSidedNoise}
\end{eqnarray}
Notice that the sign convention in (\ref{sign}) follows from the implicit assumption that
R(t) is the Fourier transform of $E[n(f) n(f')]$.  In the literature 
another convention with the minus sign is also found corresponding to 
defining R(t) as the Fourier transform of $E[n(f) n^*(f')]$. 
Using the relation $h(-f)=h(f)^{*}$, we can finally write the FIM:
\begin{equation}\label{FisherFreq}
 i_{a\,b} = E[l_a\, l_b] = \langle h_a(f)\,,\,h_b(f)\rangle
\end{equation}
where $h_a(f)$ are the derivatives of the Fourier transform of the signal with respect to the a-th parameter. 
We have introduced a mean in the frequency space:
\begin{equation}\label{MeanFreq}
\langle u(f)\,,\,v(f) \rangle \equiv 4 \mathcal{R}\bigg[ \int_{f_{low}}^{f_{cut}}{ \ud f\, \frac{u(f) v(f)* }{S_h(f)}}\bigg]
\end{equation}
where the range of integration depends on the antenna properties and the theoretical model for the binary system.

The SNR corresponds to the optimal filter: 
\begin{equation*}
\rho^2 \equiv \langle h(f)\,,\,h(f)\rangle = 4 \int_{f_{low}}^{f_{cut}}{\ud\!f \frac{|h(f)|^2}{S_h(f)}} 
\end{equation*}

We can express in the same way all the quantities we need in order to calculate the second order variance, like scalar products of $h(f)$ derivatives. 
 \begin{eqnarray}
\upsilon_{a,b}&=& - \upsilon_{ab} = i_{ab}=\langle h_{a}\,,\,h_{b} \rangle \label{vab} \\
\upsilon_{ab\,,\,c}  &=& \langle h_{ab}\,,\,h_c\rangle\\
\upsilon_{abc} &=& -\langle h_{ab}\,,\,h_c \rangle - \langle h_{ac}\,,\,h_b \rangle-\langle h_{bc}\,,\,h_a \rangle
\end{eqnarray}
\begin{eqnarray}
\upsilon_{ab\,,\,cd}&=& \langle h_{ab}\,,\,h_{cd} \rangle  + \langle h_{a}\,,\,h_{b} \rangle \langle h_{c}\,,\,h_{d} \rangle \\
\upsilon_{abc\,,\,d} &=& \langle h_{abc}\,,\,h_d\rangle
\end{eqnarray}
\begin{eqnarray}
\upsilon_{abcd}&=& -\langle h_{ab}\,,\,h_{cd} \rangle - \langle h_{ac}\,,\,h_{bd} \rangle -\langle h_{ad}\,,\,h_{bc} \rangle -\nonumber \\ 
&-&\langle h_{abc}\,,\,h_{d} \rangle-\langle h_{abd}\,,\,h_{c} \rangle-\nonumber \\&-&\langle h_{acd}\,,\,h_{b} \rangle-\langle h_{bcd}\,,\,h_{a} \rangle\\
\upsilon_{ab\,,\,c\,,\,d}&=& \langle h_{a}\,,\,h_{b} \rangle \,\langle h_{c}\,,\,h_{d} \rangle =i_{ab}\,i_{cd} \label{vabcd}
\end{eqnarray}

If one uses these functions, the form for the second order variance, eq (\ref{VarMatrix}) can be further simplified:
\begin{widetext}
\begin{eqnarray}\label{VarMatrixSimplified}	
 C_2(\vartheta^j) &=& i^{jm}i^{jn}i^{pq}(\upsilon_{nmpq}\!+\!3 \langle h_{nq}\,,\,h_{pm}\rangle + 2\upsilon_{nmp,q}\!+\!
\upsilon_{mpq,n}) +\\
&+& i^{jm}i^{jn}i^{pz}i^{qt}\bigg( v_{npm} v_{qzt} + \frac{5}{2} v_{npq} v_{mzt} + 2 v_{qz\,,\,n}v_{mtp} +2 v_{qp,z} v_{nmt} +\nonumber \\
&+&  6 v_{mqp}v_{nt\,,\,z} + v_{pqz} v_{nt\,,\,m} + 2 v_{mq\,,\,z} v_{pt\,,\,n} +2 v_{pt\,,\,z}v_{mq\,,\,n} + v_{mz\,,\,t}v_{nq\,,\,p}\bigg)\nonumber 
\end{eqnarray}
\end{widetext}
\section{Inspiral phase waveform for binary systems}\label{WaveformSection} 

We apply the general results of the theory to the case of a binary system in the inspiral phase. Starting from the the 3.5PN phasing formula (\cite{Blanchet2002}) we write the Fourier transform of the chirp signal:
\begin{equation}\label{WaveTime}
h(t)= a(t) \left[e^{i\phi(t)}+e^{-i\phi(t)}\right] 
\end{equation}
where $\phi(f)$ is the implicit solution of the 3.5PN phasing formula, using the  {\it stationary phase approximation} (SPA) (\cite{Damour2001}, \cite{Damour2002}). The final result is:

\begin{equation}\label{Waveform}
 h(f) = \mathcal{A} f^{-\frac{7}{6}} e^{i\,\psi(f)}.
\end{equation}

The phase is given by:

\begin{equation}\label{phase}
 \psi(f)\,=\, 2 \pi f t - \phi - \frac{\pi}{4} + \frac{3}{128\, \eta\, v^5} \sum_{k=0}^N \alpha_k v^k
\end{equation}

where $t$ and $\phi$ are the arrival time and the arrival phase. 
The function $v$ can be defined either in terms of the total mass of the binary system, $M=m_1+m_2$, or in terms of the chirp mass, $\mathcal{M}= \eta^\frac{3}{5}\,M$:

\begin{equation*}
v = (\pi M f)^\frac{1}{3} = (\pi f \mathcal{M})^\frac{1}{3} \eta^{-\frac{1}{5}}
\end{equation*}

where $\eta$ is the symmetrized mass ratio $$ \eta= \frac{m_1 m_2}{M^2}.$$
The amplitude is a function of the chirp mass, the effective distance, and the orientation of the source:
$$\mathcal{A} \propto \mathcal{M}^{\frac{5}{6}} Q(angles)/D.$$
The coefficients $\alpha_k$'s with $k=0..N$ (the meaning of each terms being the $\frac{k}{2}$ PN contribution) are given by:
\begin{eqnarray*}
 \alpha_0 &=& 1,\\
 \alpha_1 &=& 0,\\
 \alpha_2 &=& \frac{20}{9}\left( \frac{743}{336} + \frac{11}{4} \eta\right),\\
\alpha_3 &=& -16\pi,	\;\; \alpha_4= 10\left(\frac{3058673}{1016064} + \frac{5429}{1008} \eta + \frac{617}{144}\eta^2\right)\\
\alpha_5 &=& \pi\left( \frac{38645}{756} + \frac{38645}{252} \log{\frac{v}{v_{lso}}} - \frac{65}{9}\eta\left[1+ 3 \log{\frac{v}{v_{lso}}}\right]\right)\end{eqnarray*}
\begin{eqnarray*}
\alpha_6 &=& \left( \frac{11583231236531}{4694215680} - \frac{640}{3}\pi^2 - \frac{6848}{21}\gamma\right) + \\
&+& \eta\left( -\frac{15335597827}{3048192} + \frac{2255}{12}\pi^2 - \frac{1760}{3}\theta + \frac{12320}{9} \lambda \right) +\\
&+& \frac{76055}{1728}\eta^2 - \frac{127825}{1296} \eta^3 - \frac{6858}{21}\log{4 v}, \\
\alpha_7&=& \pi \left( \frac{77096675}{254016} + \frac{378515}{1512} \eta - \frac{74045}{756}\eta^2\right).
\end{eqnarray*} 
where $\lambda \simeq -0.6451$, $\theta \simeq -1.28$.
$\gamma$ is the Euler constant, and $v_{lso} = (\pi M f_{lso})^\frac{1}{3}$, with $f_{lso}$ the \emph{last stable orbit} frequency for a test mass in a Scharzschild space-time of mass M:
\begin{equation}\label{LSO}
f_{lso}= (6^\frac{3}{2} \pi M)^{-1}
\end{equation}
 which will be also used as upper cutoff for the integrals (\ref{MeanFreq}) .. (\ref{vabcd}).

Given the waveform (\ref{Waveform}) one can easily calculate the Fisher information matrix, and its inverse, the CRLB.
(\ref{Waveform}) contains five unknown parameters, $(\mathcal{A},t,\phi,\mathcal{M},\eta)$ (the total mass $M$ could be used instead of the chirp mass), so that one should calculate a five dimensional square matrix.

It was already observed by (eg \cite{Cutler} ) that the errors in the distance, and consequently the amplitude $\mathcal{A}$, are uncorrelated with the errors in other parameters, ie that the Fisher information is block diagonal.
We observed that this is also the case for the waveform (\ref{Waveform}) we use here for both the FIM and the second order covariance matrix. We can therefore work in a four dimensional parameter space $(t,\phi,\mathcal{M},\eta)$.
However it is worth observing that this is not an obvious point since in general the amplitude estimation can 
be coupled with other other parameters if they enter in the signal in specific ways (see ch.3 of \cite{willsky}).\\

It is also worthwhile noticing that the SNR and the amplitude $\mathcal{A}$ are related like follows:

\begin{equation}\label{SNRtoA}
 \rho^2 \equiv \langle h(f)\,,\,h(f) \rangle = 4 \,\mathcal{A}^2 \int_{f_{low}}^{f_{cut}}{\ud f\,\frac{f^{-\frac{7}{3}}}{S_h(f)}}
\end{equation}

We perform the calculations using analytical forms of the design initial and advanced LIGO noise spectrum (\ref{IniLigo}).The initial one sided power spectral density of LIGO can be written for $f\geq f_{low}$ ($S_h(f)	=\infty, \;\;f\leq f_{low}$):
  \begin{equation}\label{IniLigo}
  S_h(f))=S_0 \left[\left(4.49 x\right)^{-56} + 0.16 x^{-4.52} + 0.52 + 0.32 x^2\right],
 \end{equation}
 
Where the lower frequency cutoff value is $f_{low}=40\mbox{Hz}$, $x\equiv \frac{f}{f_0}$, $f_0= 500 \mbox{Hz}$, and $S_0=9\times 10^{-46} \mbox{Hz}^{-1}$.

The Advanced LIGO one sided psd has the following expression, for  $f\geq f_{low}$ ($S_h(f)	=\infty, \;\;f\leq f_{low}$):
 \begin{equation}\label{AdvLigo}
 S_h(f)= S_0 \left[x^{-4.14} -5 x^{-2} + 111\frac{1-x^2+x^4/2}{1+x^2/2}\right],
 \end{equation}

\begin{figure}[htb]
\centering
\begin{tabular}{cc}
\multirow{15}{9	mm}[4mm]{\small $S_h(f)^{\frac{1}{2}}$} & \\
&\includegraphics[width=7.5cm]{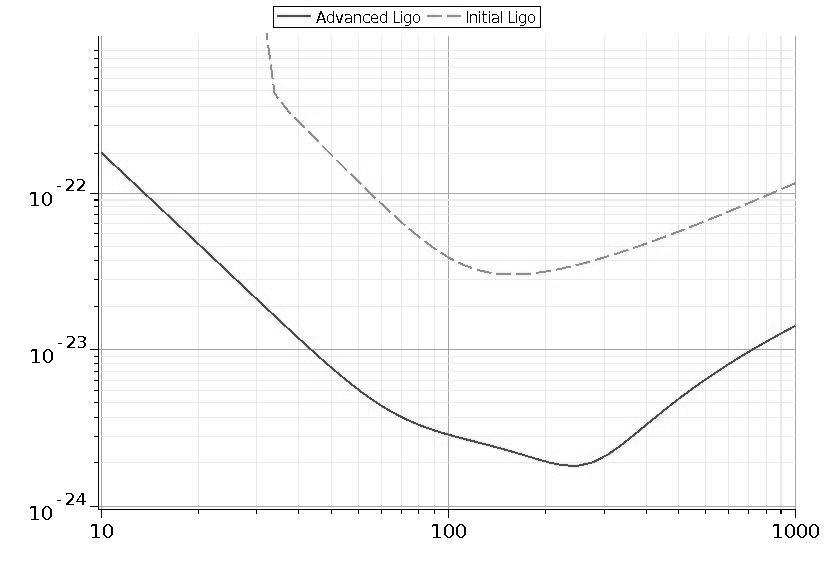}\\
& {\small $ \quad f$ }\\
\end{tabular}\hspace{-2cm} 
\caption{\small The Initial (dashed line) and Advanced (solid line) LIGO noise spectrum.}\label{LigoNoise}
\end{figure}

 Where the lower frequency cutoff value is $f_{low}=20\mbox{Hz}$, $x\equiv \frac{f}{f_0}$, $f_0= 215 \mbox{Hz}$, and $S_0= 10^{-49} \mbox{Hz}^{-1}$. They are plotted in fig. \ref{LigoNoise}

We now calculate the second order covariance for the full four dimensional parameter space $(t,\phi,\mathcal{M},\eta)$.

In order to make easier the comparison of our results with the literature we study a binary neutron star system (NS-NS), a neutron star - black hole system (NS-BH), and a black hole system (BH-BH).
A neutron star is assumed to have a mass of $1.4 M_{\odot}$ and a black hole of $10 M_{\odot}$

We performed our calculations following these steps:
(I) We give to the total mass $M$ and the mass ratio $\eta$ a numerical value depending on the binary system we consider. This makes the upper cutoff (\ref{LSO}) to have a numerical value.
(II) We compute analytically the derivatives of the wave function $h(f)$ and use it to compute the Fisher information (\ref{FisherFreq}) and his inverse, the CRLB.
(III) We calculate the $v$'s functions (\ref{vab}) .. (\ref{vabcd}) and use them to compute the second order covariance (\ref{VarMatrixSimplified})
(IV) We plot for the four parameters the CRLB, the second order variance and the total error (without bias, see end of section \ref{OneDim})

\section{One dimensional parameter space}\label{OneDim}
In this section we describe the results for the instance
where only one of the parameters (we call it $\theta$) is considered unknown, while the others have known values.
It can be shown that in this case the second order variance (\ref{VarMatrixSimplified}) can be written as:

\begin{equation}
 C_2(\theta) = \left( i^{\theta\theta}\right)^3 \left( 8\, i^{\theta \theta} \langle h_{\theta \theta}\,,\,h_{\theta}\rangle^2 - \langle h_{\theta \theta\theta}\,,\, h_{\theta}\rangle \right)
\end{equation}

Let us consider the case were $\theta$ is the arrival time $t$ in the waveform (\ref{Waveform}). 
The ratio between the second order and the first order variance turns to be:

\begin{equation}\label{OneParTime}
 \frac{C_2(t)}{\sigma_t^2}= \frac{1}{4 \rho^2}\frac{\frac{K_4}{K_0}}{\left(\frac{K_2}{K_0}\right)^2}
\end{equation}

where $K_\alpha \equiv \int{\ud f\, \frac{f^\alpha}{S_h(f)} |h(f)|^2 }$ is the $\alpha$-th moment of the signal's frequency spectrum.
The second order is negligible with respect to the first if 

\begin{equation}
 \rho^2 \gg \frac{\frac{K_4}{K_0}}{\left(\frac{K_2}{K_0}\right)^2}\,,
\end{equation}
that is, if the SNR is much larger than the {\it Kurtosis} of the signal's spectrum.
This means that for two signals with the same bandwidth, the one with a more {\it peaked} spectrum becomes requires higher SNR to attain the CRLB.
It must be noted that the $K_\alpha$ are functions of the total mass via the upper limit of the integral (see (\ref{LSO}))
\\

It can be shown that for a BH-BH (NS-NS) system the second order becomes equal to the first for $\rho=1.32$ ($\rho=1.97$). These values of SNR are smaller than those we will derive with the full parameters space. This indicates that 
it is much harder to estimate all the parameters at the same time.
Also notice that if someone requires the second order to be much smaller than the 
first, for example $10\%$ the conditions become more stringent.

\begin{figure}[htb]
 \begin{tabular}{ll}
\multirow{1}{7mm}[35mm]{\small $$\frac{M}{M_{\odot}}$$} & \includegraphics[width=6.2cm]{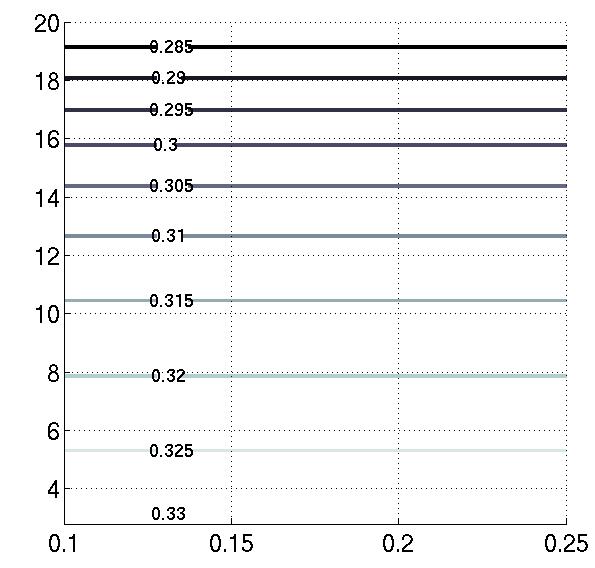}\\
\multirow{1}{7mm}[35mm]{\small $$\frac{M}{M_{\odot}}$$} & \includegraphics[width=6.2cm]{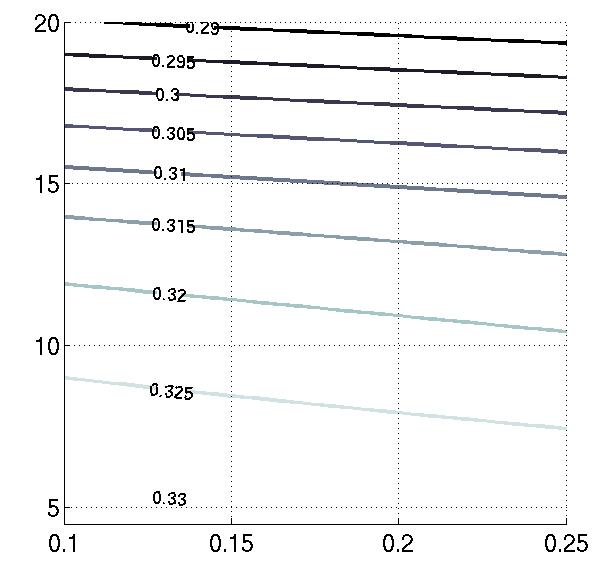}\\
\hskip 10mm& \hskip 3.0cm {\small $\eta $}  \\
\end{tabular}
\caption{The ratio $C_2(\eta)/i^{\eta\eta}$ (top) and $C_2(M)/i^{MM}$ (bottom)} \label{Ratios}
 \end{figure}

A similar calculation can be done when one of the other parameters is considered as unknown. 
For $M$ and $\eta$ the same analysis does not give a result that is equally compact and we 
only show a plot of the ratio between the second order and the first order variance, for a fixed value of SNR, $\rho=10$;
These values, are presented in Fig. \ref{Ratios}, for different values of $\eta$ and $M$ and 1PN waveforms.
When $\eta$ is the unknown parameter (upper plot), the ratio becomes smaller when the total mass of the system increases.
This is in agreement with the Monte Carlo simulations (see e.g. \cite{Cokelaer}, and section 8.3) 
performed in the full four parameter space.
The necessary SNR is not very sensitive to the value of $\eta$.
Similar considerations can be drawn when the total mass $M$ is the unknown parameter (bottom plot), except that now the 
necessary SNR seems to be slightly more sensitive to the value of $\eta$.
In both panels \ref{Ratios} the second order is always about $30\%$ of the first order.
%
\
If one works with a one dimensional parameter space more compact expressions can also be given for the first and second order bias, eqs. (\ref{BiasOne}) and (\ref{BiasTwo}):
\begin{eqnarray}
 b[1] &=& -\frac{1}{2} \left(i^{\theta\theta}\right)^2 \langle h_{\theta\theta}\,,\,h_{\theta}\rangle \\ 
b[2]&=& - (i^{\theta\theta})^3 \bigg( \frac{1}{8} \langle h_{\theta\theta\theta\theta}\,,\,h_{\theta}\rangle + \frac{5}{4} \langle h_{\theta\theta\theta}\,,\,h_{\theta\theta}\rangle -\\
&-& \frac{3}{2} \langle h_{\theta\theta\theta}\,,\, h_{\theta} \rangle - i^{\theta\theta} \langle h_{\theta\theta\theta}\,,\, h_{\theta}\rangle \langle h_{\theta\theta}\,,\, h_{\theta}\rangle -\nonumber\\
&-& \frac{9}{2} i^{\theta\theta} \langle h_{\theta\theta}\,,\, h_{\theta\theta}\rangle \langle h_{\theta\theta}\,,\, h_{\theta}\rangle + \frac{9}{8} \left(i^{\theta\theta}\right)^2 \langle h_{\theta\theta}\,,\, h_{\theta}\rangle ^3\bigg)\nonumber
\end{eqnarray}

We observed that the first two orders of the bias are, for all the the examples of this paper few order of magnitude smaller 
than the variance contributions. Therefore we do not include them in the presented results.
Ongoing research on waveforms including the merger and ringdown phases show that the bias
can become important too for systems with masses beyond the range considered here.

\section{Full parameter space}\label{fullparspace}
We present results for the full parameter space beginning with the examples obtained using the initial LIGO noise (\ref{IniLigo}) and then show 
the same results for the advanced LIGO noise (\ref{AdvLigo}). 
In each plot we show three curves as a function of the SNR.
The dotted one is the CRLB (square root of the diagonal elements of the inverse of the FIM).
The dashed one is the square root of the corresponding diagonal elements in the second order covariance matrix, and the continuous one is the square root of the sum of the diagonal elements of the
of the FIM and of the second order covariance matrix. For all the cases analyzed in this paper
the bias resulted negligible and is not included in the plots. 

For the bottom two of the
 four plots panels, the curves are divided by the actual value of the parameter in order to
express the relative uncertainty.
The general trend is that the CRLB fails to describe correctly the error for SNRs lower than 
20.  For the $t$ and the the $M$ this regime is of physical interest while for the 
symmetrized mass ratio and the phase the CRLB already predicts very large uncertainties.
It is also worth noticing that the SNR at which the second order covariance matrix 
contribution becomes comparable to the CRLB is much larger when the full parameter space 
is involved. For t for example in the NS-NS case the two are the same at $\rho=2$ while 
for the full parameter space they equate around $\rho=15$. This results appear to indicate 
that also the timing based directional reconstruction accuracy, is worse when 
the other physical parameters of the waveform are unknown.

\begin{figure}[htb]
\centering
\begin{tabular}{lc}
 \multirow{1}{9mm}[20mm]{\small $ \Delta t [\mbox{ms}]$} & \includegraphics[width=6cm,height=4.4cm]{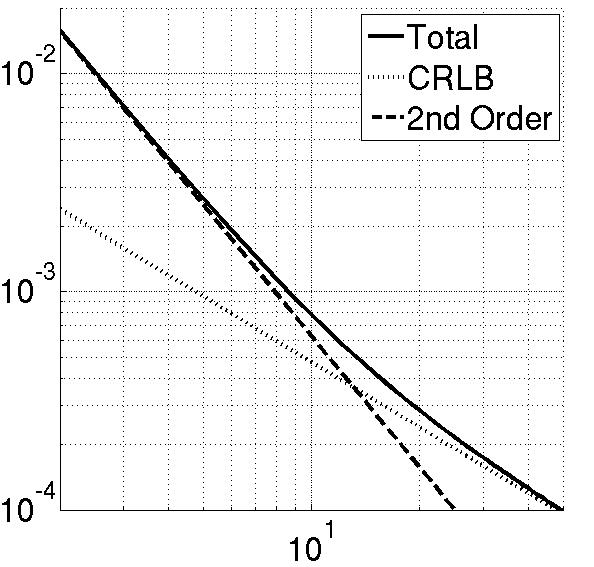} \\
 \multirow{1}{10mm}[20mm]{\small $\Delta \phi$[rad]} &  \includegraphics[width=6cm,height=4.4cm]{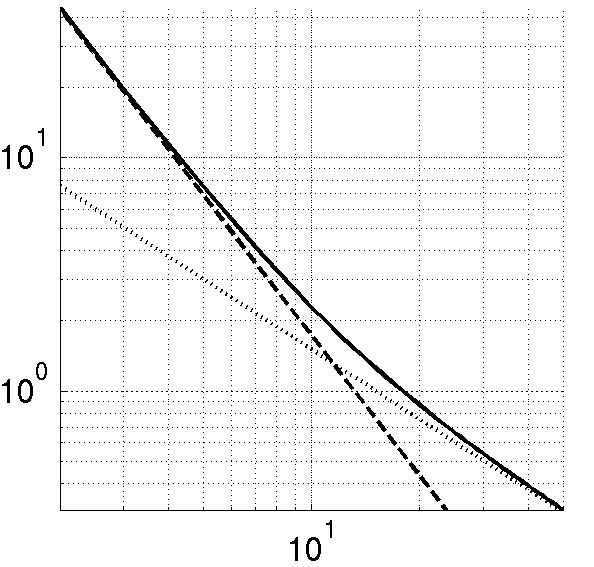}\\
  \multirow{1}{9mm}[20mm]{ $ \frac{\Delta \mathcal{M}}{\mathcal{M}} [\%]$} & \includegraphics[width=6cm,height=4.4cm]{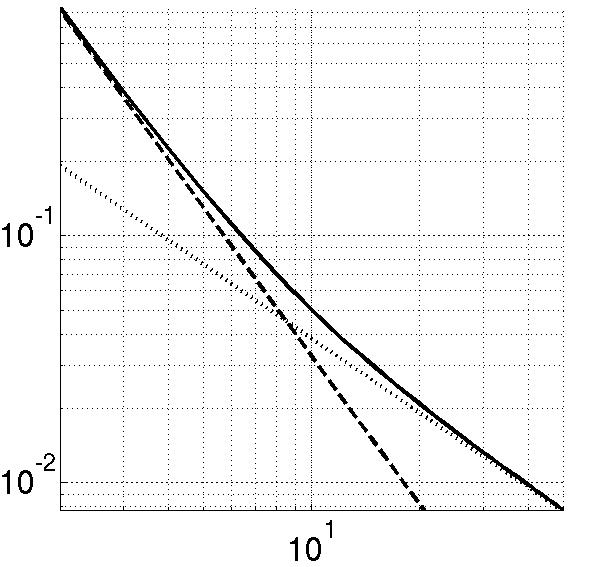} \\
 \multirow{1}{9mm}[20mm]{ $ \frac{\Delta \eta}{\eta} [\%]$} & \includegraphics[width=6cm,height=4.4cm]{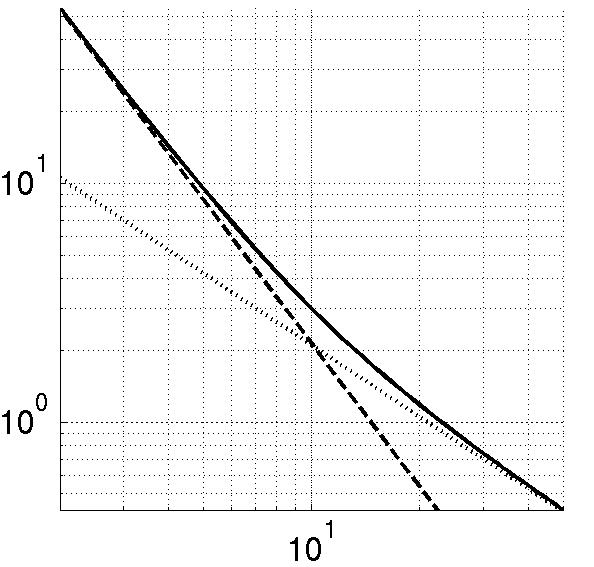}\\ 
&{ {\small $\rho$}} 
\end{tabular}
 \caption{{\small  NS-NS signal in initial LIGO noise . 
 The dotted line is the CRLB.
 The dashed line is the square root of the second order covariance matrix, 
 and the continuous line is the square root of the sum of the diagonal elements of the FIM and of the second order covariance matrix.
  In the last two panels the errors are divided by the value of the parameter.}}\label{35NSNSINIL}
\end{figure}

\begin{figure}[htb]
 \begin{tabular}{lc}
\multirow{1}{9mm}[20mm]{\small $ \Delta t [\mbox{ms}]$} & 
\includegraphics[width=6cm,height=4.4cm]{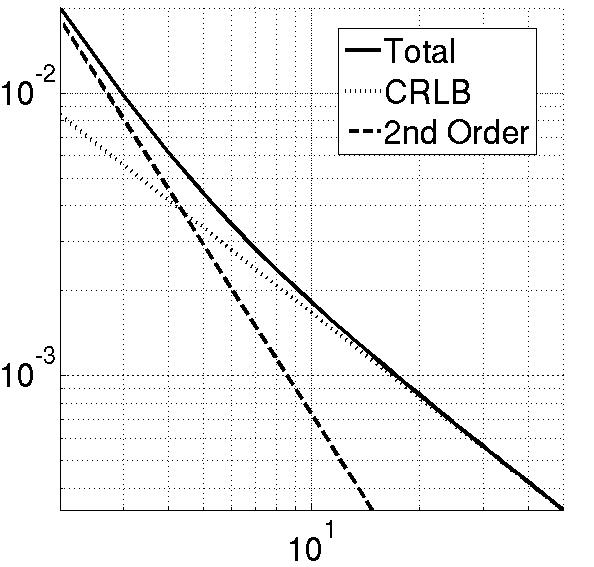} \\
 \multirow{1}{10mm}[20mm]{\small $\Delta \phi$[rad]} &
 \includegraphics[width=6cm,height=4.4cm]{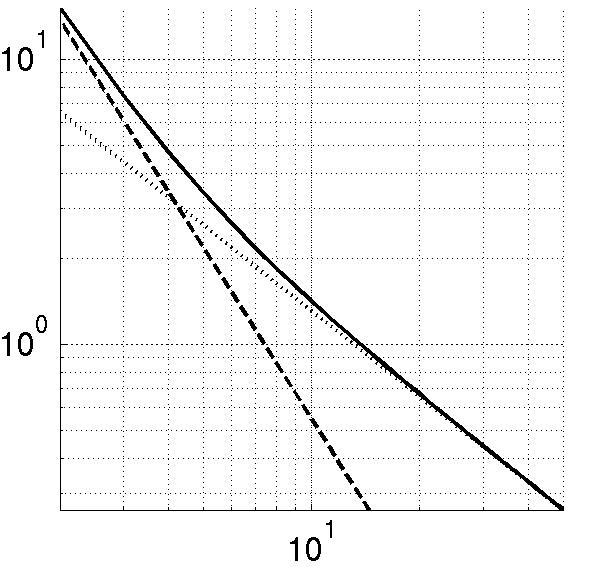}\\
 \multirow{1}{9mm}[20mm]{ $ \frac{\Delta \mathcal{M}}{\mathcal{M}} [\%]$} & \includegraphics[width=6cm,height=4.4cm]{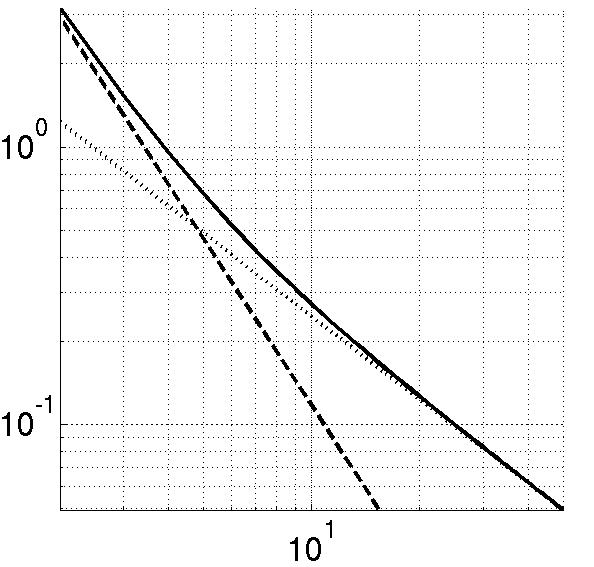} \\
 \multirow{1}{9mm}[20mm]{ $ \frac{\Delta \eta}{\eta} [\%]$} & \includegraphics[width=6cm,height=4.4cm]{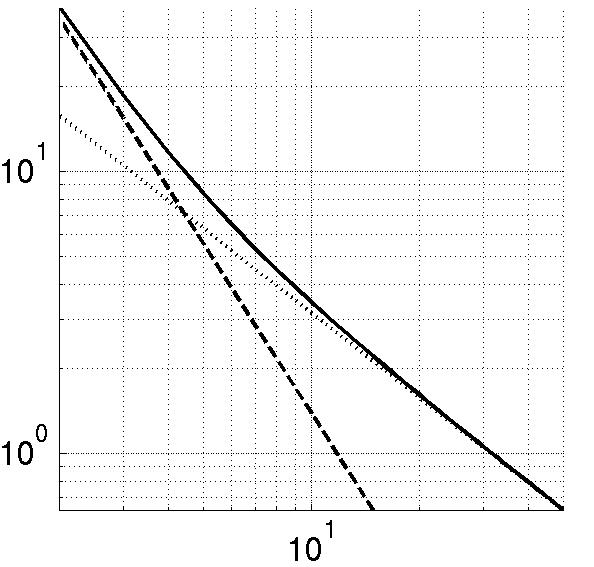}\\ 
&{ {\small $\rho$}}  
\end{tabular}
\caption{{\small NS-BH signal in initial LIGO noise.
The dotted line is the CRLB.
 The dashed line is the square root of the second order covariance matrix, 
 and the continuous line is the square root of the sum of the diagonal elements of the FIM and of the second order covariance matrix. In the last two panels the errors are divided by the value of the parameter.}}
\end{figure}

\begin{figure}[htb]
\begin{tabular}{lc}
\multirow{1}{9mm}[20mm]{\small $ \Delta t [\mbox{ms}]$} & 
\includegraphics[width=6cm,height=4.4cm]{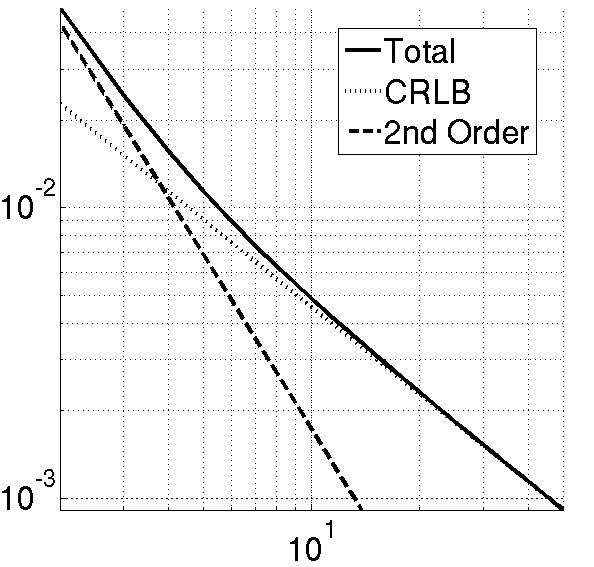} \\
 \multirow{1}{10mm}[20mm]{\small $\Delta \phi$[rad]} &
 \includegraphics[width=6cm,height=4.4cm]{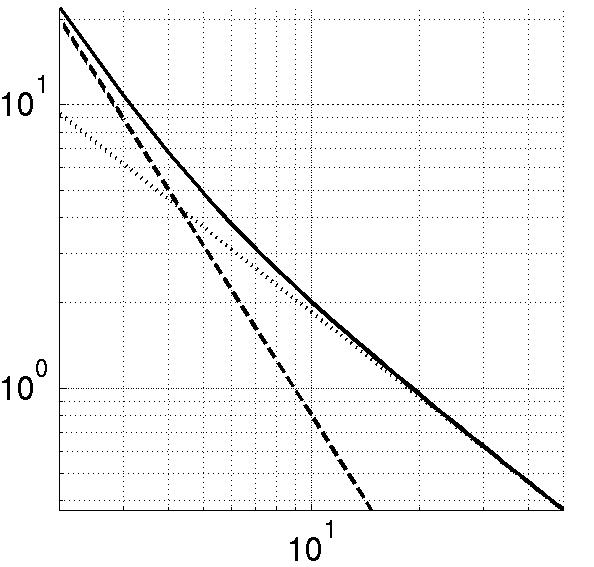}\\
 \multirow{1}{9mm}[20mm]{ $ \frac{\Delta \mathcal{M}}{\mathcal{M}} [\%]$} & \includegraphics[width=6cm,height=4.4cm]{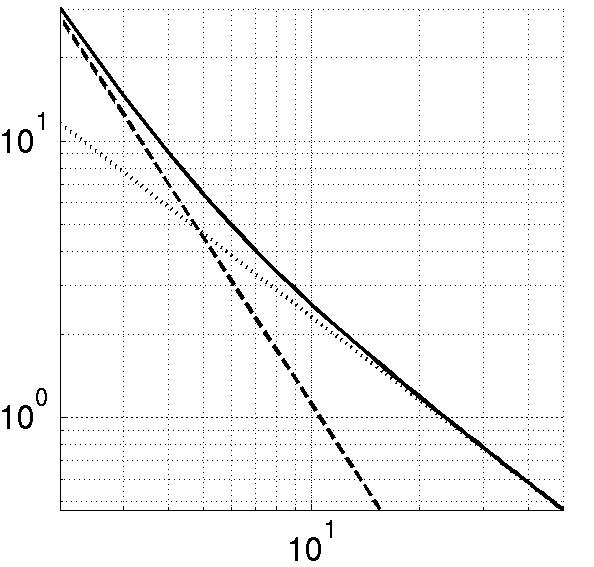} \\
 \multirow{1}{9mm}[20mm]{ $ \frac{\Delta \eta}{\eta} [\%]$} & \includegraphics[width=6cm,height=4.4cm]{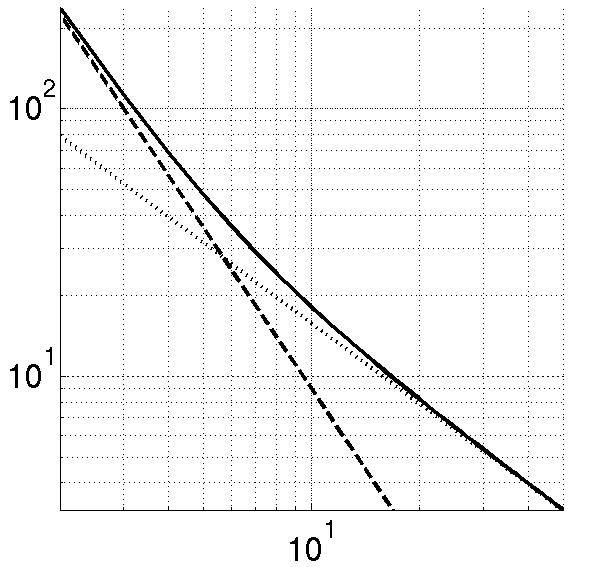}\\ 
&{ {\small $\rho$}} 
\end{tabular}
 \caption{{\small BH-BH signal in initial LIGO noise.The dotted line is the CRLB.
 The dashed line is the square root of the second order covariance matrix, 
 and the continuous line is the square root of the sum of the diagonal elements of the FIM and of the second order covariance matrix.
 In the last two panels the errors are divided by the value of the parameter.}}
\end{figure}
\clearpage
\begin{figure}[htb]
\centering
\begin{tabular}{lc}
\multirow{1}{9mm}[20mm]{\small $ \Delta t [\mbox{ms}]$} & 
\includegraphics[width=6cm,height=4.4cm]{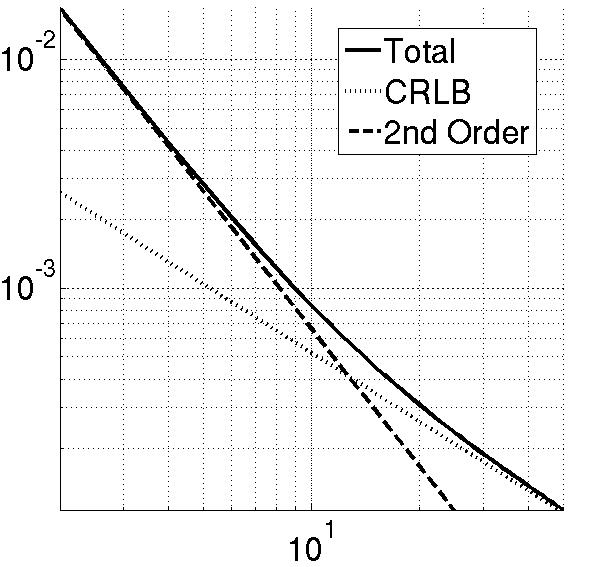} \\
 \multirow{1}{10mm}[20mm]{\small $\Delta \phi$[rad]} &
 \includegraphics[width=6cm,height=4.4cm]{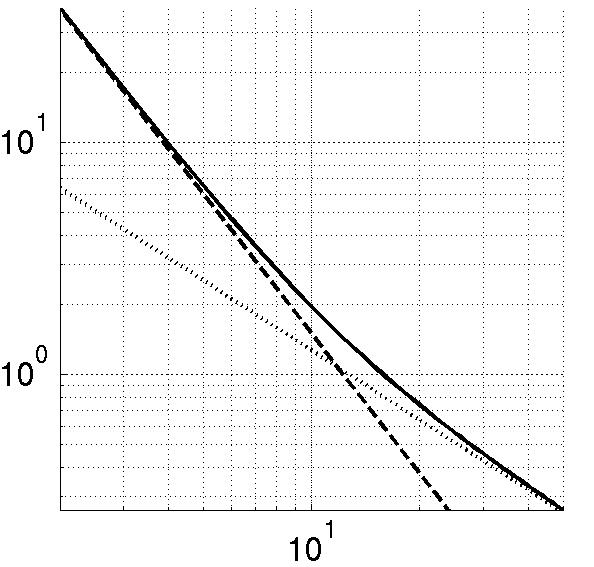}\\
 \multirow{1}{9mm}[20mm]{ $ \frac{\Delta \mathcal{M}}{\mathcal{M}} [\%]$} & \includegraphics[width=6cm,height=4.4cm]{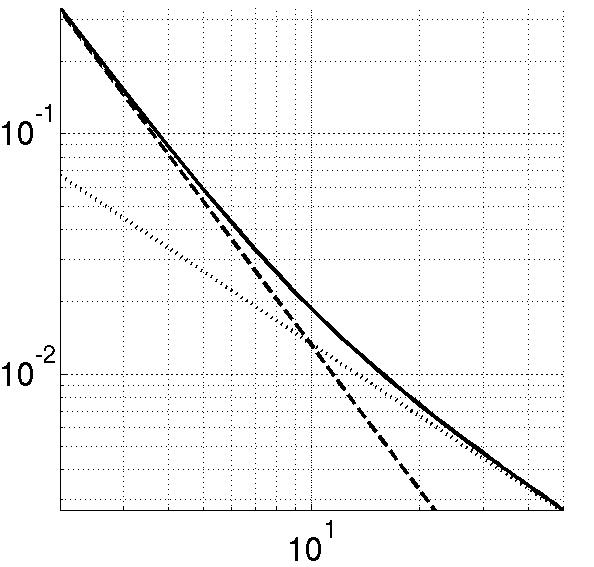} \\
 \multirow{1}{9mm}[20mm]{ $ \frac{\Delta \eta}{\eta} [\%]$} & \includegraphics[width=6cm,height=4.4cm]{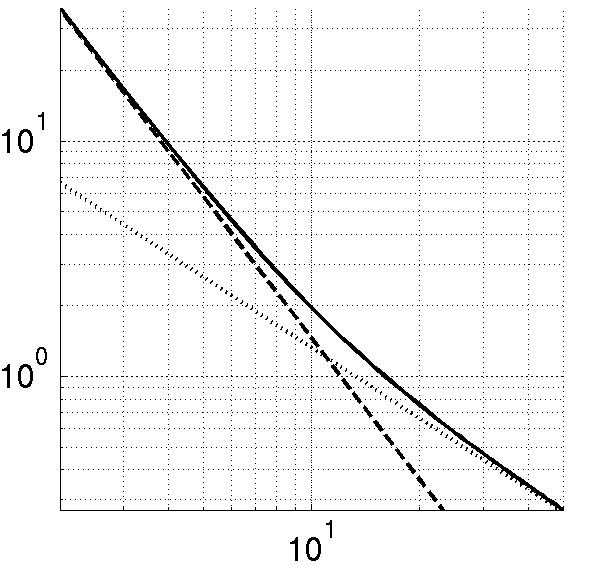}\\ 
&{ {\small $\rho$}} 
\end{tabular}\caption{{\small NS-NS signal with Advanced LIGO noise . The dotted line is the CRLB.
 The dashed line is the square root of the second order covariance matrix, 
 and the continuous line is the square root of the sum of the diagonal elements of the FIM and of the second order covariance matrix.
 In the last two panels the errors are divided by the value of the parameter. }}\label{35NSNSADVL}
\end{figure}

\begin{figure}[htb]
\centering
\begin{tabular}{lc}
\multirow{1}{9mm}[20mm]{\small $ \Delta t [\mbox{ms}]$} & 
\includegraphics[width=6cm,height=4.4cm]{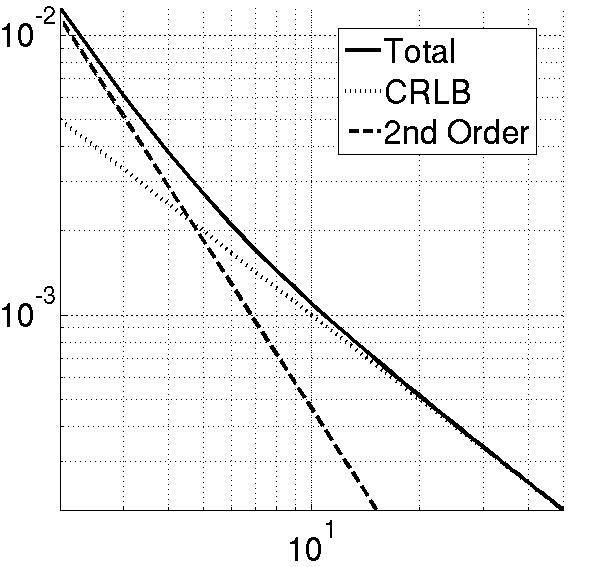} \\
\multirow{1}{10mm}[20mm]{\small $\Delta \phi$[rad]} & 
\includegraphics[width=6cm,height=4.4cm]{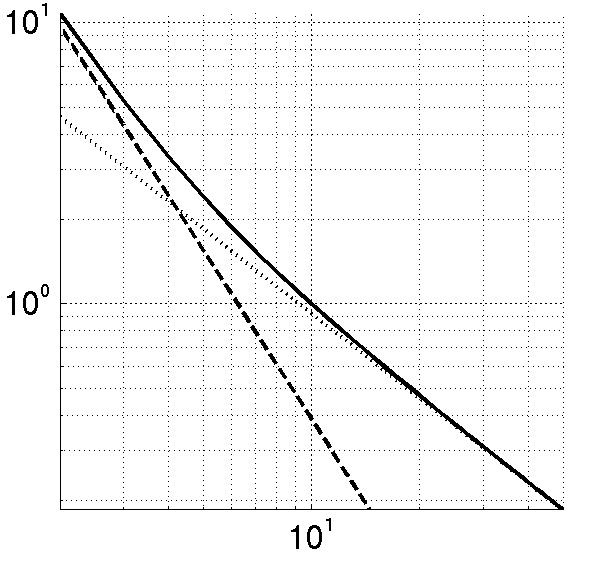}\\
\multirow{1}{9mm}[20mm]{ $ \frac{\Delta \mathcal{M}}{\mathcal{M}} [\%]$} &
 \includegraphics[width=6cm,height=4.4cm]{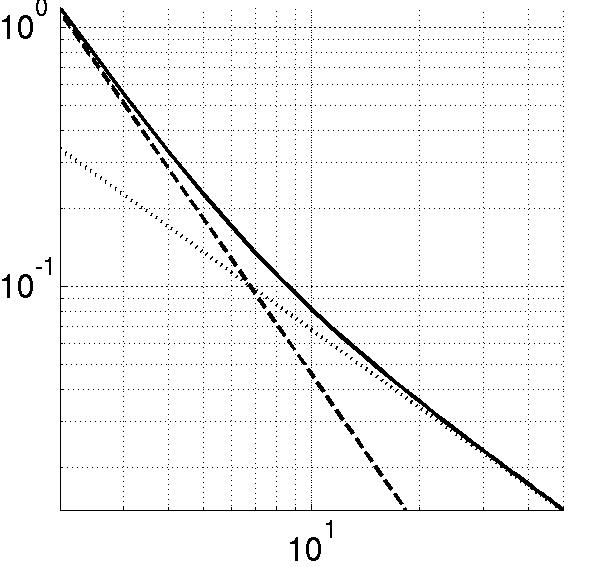} \\
 \multirow{1}{9mm}[20mm]{ $ \frac{\Delta \eta}{\eta} [\%]$} & \includegraphics[width=6cm,height=4.4cm]{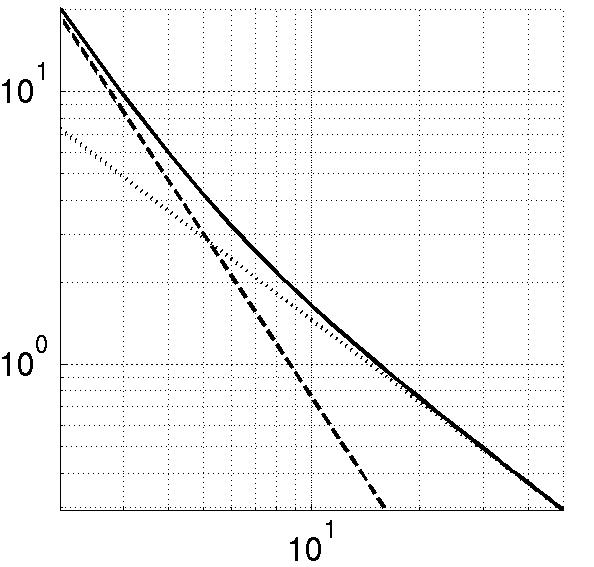}\\ 
&{ {\small $\rho$}} 
\end{tabular}
 \caption{{\small  NS-BH signal with advanced LIGO noise. The dotted line is the CRLB.
 The dashed line is the square root of the second order covariance matrix, 
 and the continuous line is the square root of the sum of the diagonal elements of the FIM and of the second order covariance matrix.
 In the last two panels the errors are divided by the value of the parameter. }}
\end{figure}

\begin{figure}[htb]
\centering
\begin{tabular}{lc}
\multirow{1}{9mm}[20mm]{\small $ \Delta t [\mbox{ms}]$} & 
\includegraphics[width=6cm,height=4.4cm]{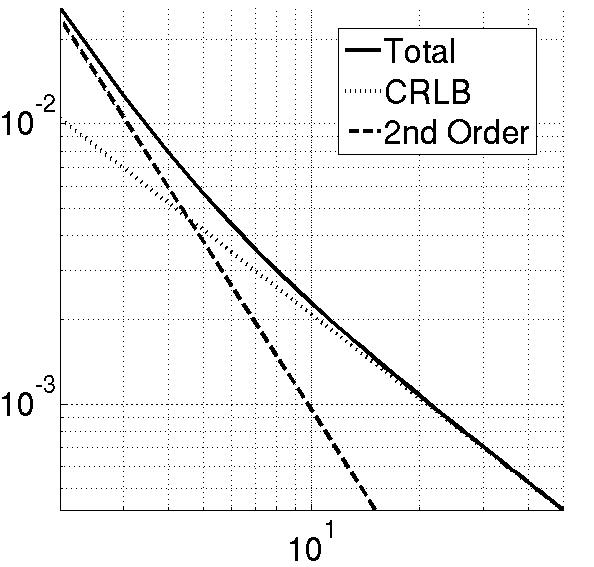} \\
\multirow{1}{10mm}[20mm]{\small $\Delta \phi$[rad]} & 
\includegraphics[width=6cm,height=4.4cm]{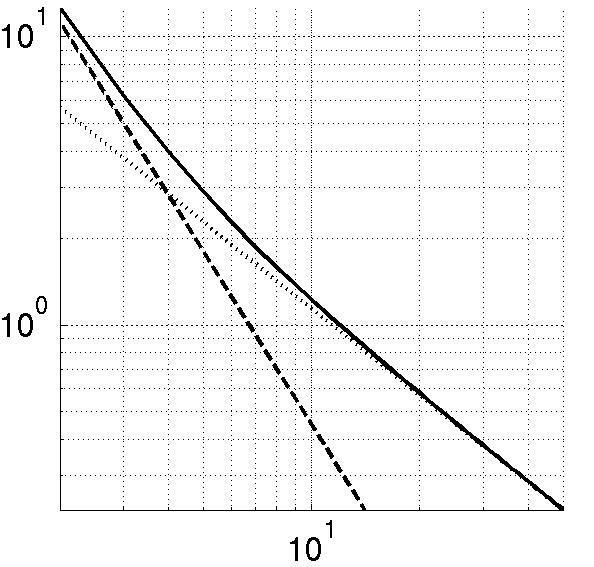}\\
\multirow{1}{9mm}[20mm]{ $ \frac{\Delta \mathcal{M}}{\mathcal{M}} [\%]$} &
 \includegraphics[width=6cm,height=4.4cm]{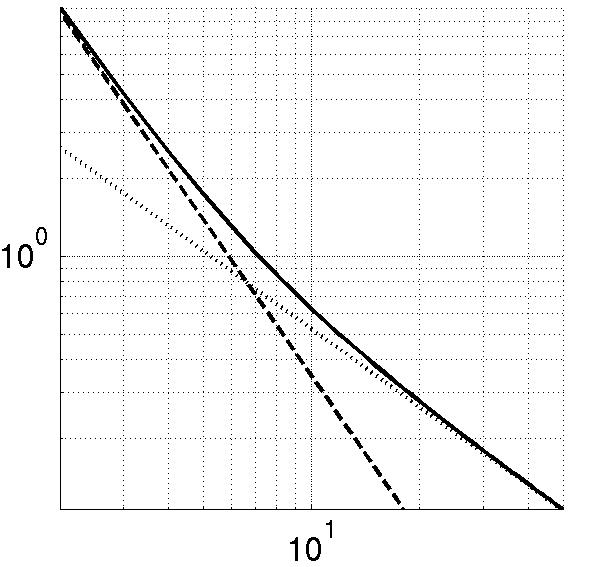} \\
 \multirow{1}{9mm}[20mm]{ $ \frac{\Delta \eta}{\eta} [\%]$} & \includegraphics[width=6cm,height=4.4cm]{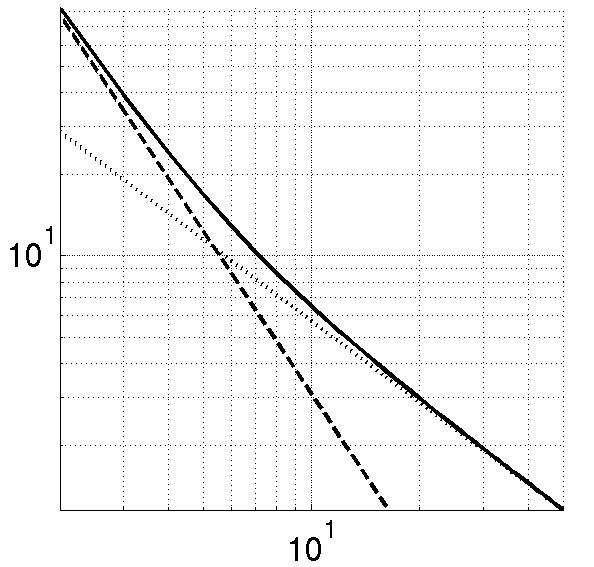}\\ 
&{ {\small $\rho$}} 
\end{tabular}
\caption{{\small BH-BH signal with advanced LIGO noise. The dotted line is the CRLB.
 The dashed line is the square root of the second order covariance matrix, 
 and the continuous line is the square root of the sum of the diagonal elements of the FIM and of the second order covariance matrix.
 In the last two panels the errors are divided by the value of the parameter.}}
\end{figure}
\begin{figure}[htb]
\begin{tabular}{|lc|} 
\hline
& NS-NS\\
& \includegraphics[width=6.3cm,height=4.6cm]{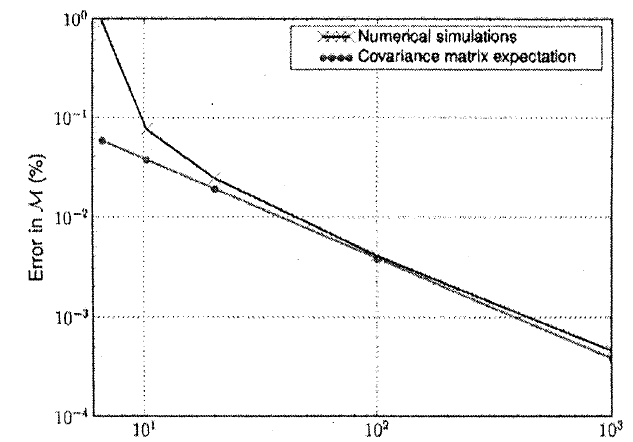}\\
\multirow{1}{7mm}[20mm]{ $ \frac{\Delta \mathcal{M}}{\mathcal{M}} [\%]$} &
 \;\includegraphics[width=6.3cm,height=4.6cm]{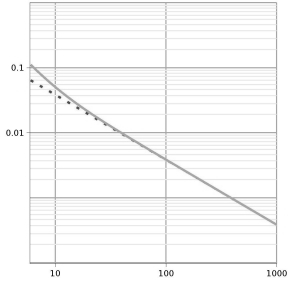} \\
\hline & low BH-BH\\ 
&\includegraphics[width=6.3cm,height=4.6cm]{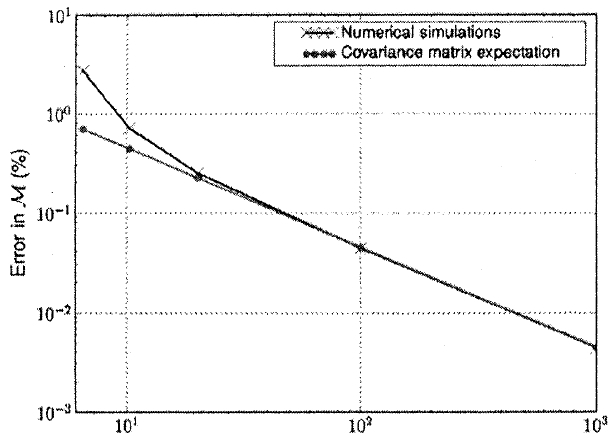}\\
\multirow{1}{9mm}[20mm]{$ \frac{\Delta \mathcal{M}}{\mathcal{M}} [\%]$} & \includegraphics[width=6.3cm,height=4.6cm]{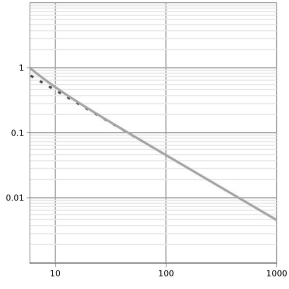}\\ 
 &{ {\small $\rho$}} \\
 \hline
\end{tabular}
\caption{{\small The simulations of T. Cokelaer (first and thrid panel from top), compared with our result (see the text for a discussion).}}\label{vscokelaer}
\end{figure}\clearpage
It interesting to compare our results with Monte Carlo simulations for 3.5PN waveforms, (\cite{Cokelaer}).

 In fig. \ref{vscokelaer} we reproduce the plots of (\cite{Cokelaer}) obtained for a NS-NS system (first panel from top) and for a low mass BH-BH system  - $10 M_{\odot}$ - (third panel from to) when the initial LIGO noise is used. 

The results we obtain for the same systems are shown in the second and fourth panels from the top of fig. \ref{vscokelaer}, where the CRLB and the square root of the sum of the inverse FIM and the second order covariance are plotted.
The plots show that the use of the second order covariance predicts correctly the SNR regime 
where the CRLB fails to describe the MLE accuracy. 
The explicit percentage discrepancies presented for SNR equal to 20, 10 and 6 in table \ref{montecarlo} 
seem to indicate that higher orders of the expansions might be necessary to fully reproduce the 
error derived with the Monte Carlo simulations.
 
\begin{table}
\begin{tabular}{rrrr}\hline\hline
NS-NS SNR & 20 &  10  &  6    \\ \hline
 Cokelaer   &  $25\%$ &  $200\%$  &  $700\%$ \\
 Z.V.M.   &    $8.6\%$ &  $131\%$  &  $174\%$  \\ \hline
BH-BH SNR & 20 &  10  &  6    \\ \hline
 Cokelaer   &  $10\%$ &  $150\%$  &  $230\%$ \\  
 Z.V.M.   &     $2.9\%$ &  $111\%$  &  $129\%$ \\
                \hline\hline
                \end{tabular}
        \caption{ The discrepancies between the CRLB error prediction and Montecarlo simulations
are presented above the discrepancies observed using the first and second order covariance matrix. The discrepancies are presented 
for three values of the SNR: 20,10 and 6.}
        \label{montecarlo}
\end{table}

\section{Conclusions}\label{conclusions}
In this paper we applied a recently derived statistical methodology to
gravitational waves generated by the inspiral phase of binary mergers 
and for noise spectral densities of gravitational wave interferometers.
Explicitly 
we computed the first two orders 
of MLE expansions of bias and covariance matrix to evaluate MLE uncertainties.
We also compared the improved error estimate with existing numerical estimates.
The value of the second order of the variance expansions allows to
get error predictions closer to what is observed in numerical simulations
than the inverse of the FIM.
The condition where the second order covariance is negligible with respect to the first order 
predicts correctly the necessary SNR to approximate the error
with the CRLB and provides new insight on the
relationship between waveform properties SNRs and estimation errors.
Future applications include IMR waveforms, network of detectors and source location estimation.

\section{Acknowledgments}
MZ would like to thank the National Science Foundation for the support through the awards
NSF0855567 and NSF0919034, while SV would like to thank the Universite' Pierre-et-Marie-Curie and 
Embry Riddle Aeronautical university for logistic and financial support.

\section{Appendix A: Derivation of the Expansions}\label{LSA}
Analytic expressions for the moments of a (MLE) are often difficult
to obtain given a non-linear data model.
However, it is known from (\cite{Haldane}) that likelihood expansions can be used to obtain approximate expressions for
the moments of a MLE in terms of series expansions in inverse powers of the sample size
($n$). These expansions are valid in small samples or SNR where the MLE may have
significant bias and may not attain minimum variance.

An expressions for the second covariance equivalent to the one used here, was first given in a longer form 
(about a factor 2), and with the prescription of isolating the different asymptotic orders inside the tensors, in (\cite{Naftali2001}). The expression presented here were first derived in an unpublished MIT technical report (\cite{ZanolinMIT}) by two of the authors of this paper, where (a) the Bartlett identities (\cite{Bartlett1953},\cite{Bartlett1953-}) were used to simplify the expression of the second order variance,  and derive the second order bias. (b) the prescription on the tensors no longer needed to be implemented. The final form of the second order covariance has already been published 
without proof, in two papers (\cite{Zanolin2001}, \cite{Zanolin2004}), were the first and third author of this paper are coauthors, involving MLE of source and environment parameters that use acoustic propagation within shallow water ocean wave guide.

In this section, we derive the second order bias for a multivariate MLE and we introduce a chain rule that allows the derivation of the of the second order covariance matrix from the second order bias.
The explanation follows closely ~\cite{ZanolinMIT}. 
The derivation of the bias is performed in two steps: first we derive the expansions in the non physical scenario
of n statistically independent identical measurements and then set n=1 for the case of interest of this paper.\\

The derivation follows the approach of (\cite{Lawley}) for likelihood expansions, 
(\cite{McCullagh1987}) for the large sample approximations and (\cite{BarnNiel1986}, \cite{BarnNiel1999})
for the notation of the observed likelihood expansions.
For the small sample case of interest, $asymptotic$ orders generated
from the likelihood expansion may contain different powers of $\frac{1}{n}$,
and the contributions to a single power of  $\frac{1}{n}$
may have to be collected from different $asymptotic$ orders.
(\cite{Shenton1963}) extending the work of (\cite{Haldane}), confronted the equivalent problem
of separating the powers in $n$ within the expectations of products of linear forms and arrived at expressions for the second order
bias of the MLE by embedding the derivation in a discrete probability scheme.
Some applications of their results are given in
(\cite{Bowman1983}), (\cite{Shenton1986}), (\cite{Bowman1992}),(\cite{Bowman1998}),(\cite{Shenton1990}).
We present here an independent derivation for the second order bias that is valid for general discrete or continuous random variables.\\
Let us consider a set of $n$ independent and identically distributed 
experimental 
data vectors ${\underline{x}}_i=\{x_{i1},..,x_{iN} \}$,
where $N$ is the dimension of every vector.
We assume that these data are described by a probability density 
$P(\underline{{x}},\underline{\vartheta})=\Pi_ip({\underline{x}}_i,\underline{\vartheta})$ that depends 
on a D-dimensional parameter vector $ \underline{{\vartheta}}=\{
{{\vartheta}}_1,..,{{\vartheta}}_D \}  $.
According to (\cite{Lawley}), we suppose that the MLE $ \underline{\widehat{\vartheta}}=\{
{\widehat{\vartheta}}_1,..,{\widehat{\vartheta}}_D \}  $
of  $\underline{\vartheta}$ is given by a stationary point of the  
likelihood function $l(\underline{x},\underline{\vartheta}) 
=ln(P(\underline{x},\underline{\vartheta}))=\sum_{i=1}^{n}ln(p(\underline{x}_i,\underline{\vartheta}))$ with respect to the components of  
$\underline{\vartheta}$ 
\begin{eqnarray}
  l_r(\underline{x},\underline{\widehat{\vartheta}}) =\label{stazionarieta} \frac{\partial l(\underline{x},\underline{\vartheta})}{\partial\vartheta_r }{{\mid}_{\underline{\vartheta}=\underline{\widehat{\vartheta}}}}=0 \,\,\,\,\,\,\, r=1,..,D.
\end{eqnarray}
The first step in deriving the likelihood expansion, 
if $l_r(\underline{x},\underline{\widehat{\vartheta}})$ can be expanded
as a Taylor series in the components of the observed error 
$\underline{\widehat{\vartheta }}-\underline{\vartheta}$,
consists of writing 
 $l_r(\underline{x},\underline{\widehat{\vartheta}})$ as
\begin{widetext}
\begin{eqnarray}
0\!=\!l_r(\underline{x},\underline{\widehat{\vartheta}})\!=\!l_r(\underline{x},(\underline{\widehat{\vartheta}}\!-\!\underline{\vartheta})\!+\!\underline{\vartheta})\!=\!l_r(\underline{x},\underline{\vartheta})\!+\! \label{cippo} l_{rs}(\underline{x},\underline{\vartheta})( \widehat{\vartheta }\!\!-\!\vartheta )^s\!
+\!\frac{1}{2}l_{rtu}(\underline{x},\underline{\vartheta})( \widehat{\vartheta }\!-\!\vartheta )^t 
( \widehat{\vartheta }\!\!-\!\vartheta )^u \!+\!...
\end{eqnarray}
\end{widetext}
where   $( \widehat{\vartheta }-\vartheta)^r $ for
$r=1,..,D$ are the components of the observed error. 
We will use the notation 
\begin{eqnarray}
\upsilon_{\{a_1 a_2..a_s\}\,\, ..\,\,\{ b_1 b_2..b_s\}}&=&
E[H_{a_1 a_2.. a_s} \,\, ..\, \,H_{b_1 b_2  ..  b_s}]  \nonumber 
\end{eqnarray}
where  $H_{a_1 a_2\,.. \,a_s}= l_{a_1 a_2\,.. \,a_s}-\upsilon_{a_1 a_2\,.. \,a_s}$
and $-\upsilon_{ab}$ is the information matrix
 $i_{ab} =-\upsilon_{ab}= -E[l_{ab}]=E[l_{a}l_{b}]$.
Introducing $j^{ab}$  as the inverse of the matrix whose elements are given by 
$j_{ab}= -l_{ab}$, equation (\ref{cippo}) can be rearranged to solve for 
$( \widehat{\vartheta }_r-\vartheta_r )=( \widehat{\vartheta }-\vartheta )^r$  
\begin{widetext}
\begin{eqnarray}
( \widehat{\vartheta }-\vartheta )^r&=&j^{rs}l_{s}+\frac{1}{2}j^{rs}l_{stu}
( \widehat{\vartheta }-\vartheta )^t ( \widehat{\vartheta }-\vartheta )^u 
+\frac{1}{6}j^{rs}l_{stuv}( \widehat{\vartheta }-\vartheta )^t
( \widehat{\vartheta }-\vartheta)^u ( \widehat{\vartheta }-\vartheta )^v \nonumber \\ 
\,\,\,\,\,&+& \frac{1}{24}j^{rs}l_{stuvw}( \widehat{\vartheta }-\vartheta )
^t\label{2}( \widehat{\vartheta }-\vartheta )^u( \widehat{\vartheta }-\vartheta )^v 
( \widehat{\vartheta }-\vartheta )^w+\ldots\,\,\,\,\,.
\end{eqnarray}
\end{widetext}
Finally we can iterate  equation (\ref{2}) with respect to the components of the observed error 
and expand  $j^{ab}$ in terms of
the information matrix inverse 
$i^{ab}=(i^{-1})_{ab}$ (\cite{BarnNiel1994}, page 149)
\begin{eqnarray}
[j^{-1}]_{ab}&=&j^{ab}=\label{barnone}[(1-i^{-1}(i-j))^{-1}i^{-1}]^{ab}=\nonumber
\\ 
&=&	{i^{ab}}+{i^{at}i^{bu}H_{tu}}+{i^{at}i^{bu}i^{vw}H_{tv}H_{uw}}+\nonumber\\
&+&{i^{at}i^{bu}i^{vw}i^{cd}H_{tv}H_{wc}H_{du}}+...\,\,\,\,\,\,\,\,\,
\end{eqnarray}
From  (\ref{2}) and (\ref{barnone})
 the terms that contribute to each $asymptotic$ order of the observed error can be obtained. 
However, in order to isolate the $asymptotic$ orders necessary to compute the second order bias 
we have chosen only a finite number of terms within equations (\ref{2}) and (\ref{barnone}).
This choice 
 can be made by $(a)$ observing that
$\upsilon_{a_1 a_2..a_s}$ is proportional to $n$  
\begin{eqnarray}\label{UpsilonDef}
\upsilon_{a_1 a_2..a_s}        
   =E[l_{a_1 a_2.. a_s}] =n \, 
E[\frac{\partial^s \,ln( p(\underline{x}_i,\underline{\vartheta}) )
}{\partial\vartheta_{a_1}\partial\vartheta_{a_2}..\partial\vartheta_{a_s} } ]
\end{eqnarray}
where the value of $i$ is irrelevant because all the vector data have the same distribution, and 
$(b)$ using the large sample approximation expressed by (\cite{McCullagh1987}, page 221)
\begin{eqnarray}\label{5}
 H_{a_1 a_2..a_s}   = l_{a_1 a_2..a_s}  -
E[l_{a_1 a_2..a_s}]\sim n^{\frac{1}{2}},
\end{eqnarray}
and proved for discrete and continuous random variables in the next paragraph.

Equation (\ref{5}) indicates that the expectation of a product of $c$ generic factors 
$ H_{a_1 a_2..a_{s}}$ is a polynomial of integer powers of $n$, 
where the highest power of $n$ is the largest integer less than or equal to $\frac{c}{2}$ (we use the notation $int(\frac{c}{2})$).
For example, 
$\upsilon_{\{a_1 a_2..a_{s}\}\{ b_1 b_2..b_{p}\}\{ c_1 c_2..c_{q} \}}$
 is proportional to $n$.\\

The proof of the large sample approximation  (\ref{5}) is an extension to continuous random variables
of the analysis performed in (\cite{Shenton1977},page 36), for discrete random variables.\\     
To prove equation (\ref{5}) we show that the quantities 
\begin{eqnarray}
P_m(n)=\label{esempio}E[H_{\underline{\alpha}_{1}}H_{\underline{\alpha}_{2}}...H_{\underline{\alpha}_{m}}]
\end{eqnarray}
obtained as the expectation of products of $m$ functions
$ H_{\underline{{\alpha}}_{j}}=\sum_{i=1}^{n}h_{i,{\underline{\alpha}}_j}$ are polynomials in integer powers of $n$ of order less than  or equal to  
$\frac{m}{2}$.\\ 
Here, $h_{i,{\underline{\alpha}}_j} = l_{i,{\underline{\alpha}}_j} - <l_{i,{\underline{\alpha}}_j}>$. 
The subscripts $\underline{\alpha}_{i}$ that appear in  
$H_{\underline{\alpha}_{i}}$ represent collections of indexes as 
introduced in equation (\ref{5}).\\
The factors $H_{\underline{\alpha}_{1}}H_{\underline{\alpha}_{2}}...H_{\underline{\alpha}_{m}}$ appearing in the expectation can be seen as the first $m$ terms of a succession  $H_{\underline{\alpha}_{i}}$,
where we choose arbitrary $H_{\underline{\alpha}_{i}}$ for $i>m$.
Using such a succession  we can define the quantity 
\begin{eqnarray*}
F(y)&=& \sum_{s=0}^{\infty}y^s P_s(n)=\sum_{s=0}^{\infty} y^sE[
(\sum_{j_1=1}^{n}h_{j_1,\underline{\alpha}_{1}})...(\sum_{j_s=1}^{n}h_{j_s,\underline{\alpha}_{s}})]
\end{eqnarray*}
where we assume that all the expectations exist.

Moving all the sums outside of the expectations and redefining the notation 
in the sums, we can obtain
\begin{widetext}
 \begin{eqnarray}
F(y)=\label{100}\sum_{s=0}^{\infty} y^s \!\!\!\!\sum_{a_{1,\underline{\alpha}_{1}}+..+a_{n,\underline{\alpha}_{1}}+
..+a_{1,\underline{\alpha}_{s}}+..+a_{n,\underline{\alpha}_{s}}=s}^{a_{p,\underline{\alpha}_{q}}=0,1}
\!\!\!\!\!\!\!\!\!\!\!\!\!\!\!\!\!\!\!\!\!\!\!\!\!\!\!\!\!E[{h
_{1,\underline{\alpha}_{1}}\!\!\!\!^{a_{1,\underline{\alpha}_{1}} }}..h_{n,\underline{\alpha}_{1}}
\!\!\!\!^{a_{n,\underline{\alpha}_{1}} }..h_{1,\underline{\alpha}_{s}}
\!\!\!\!^{a_{1,\underline{\alpha}_{s}} }..h_{n,\underline{\alpha}_{s}}
\!\!\!\!^{a_{n,\underline{\alpha}_{s}} }] 
\end{eqnarray}
\end{widetext}
where the quantity $a_{p,\underline{\alpha}_{q}}$ can be equal to 0 or to 1 
to indicate that the quantity $h_{p,\underline{\alpha}_{q}}$ is present or not in the product 
inside the expectation.
The summation over all $a_{i,\underline{\alpha}_{j}}$ is intended to account for all 
possible configurations of the indexes
(choosing some equal to zero and the others equal to one)
with the constraint that the number of factors within the expectation is equal to $s$.
We can now group together terms $(h_{i,\underline{\alpha}_{j}})^{a_{i,\underline{\alpha}_j}}$ for 
different $i$ and define $\sum_{j=1}^{s}a_{i,\underline{\alpha}_j} = s_i$, where 
$\sum_{i=1}^{n}s_i = s$. In this manner we can also factorize the expectations in 
(\ref{100}) as expectations for different data vectors. 
By making use of the statistical independence of the different data vectors, and defining 
$\lambda(a_{i,\underline{\alpha}_{j_1}}, \ldots, a_{i,\underline{\alpha}_{j_p}}) = E[h_{i,\underline{\alpha}_{j_1}} \ldots h_{i,\underline{\alpha}_{j_p}}]$, (\ref{100}) can be rewritten as follows:
\begin{widetext}
\begin{eqnarray}
F(y) &=& \sum_{s = 0}^{\infty} y^s \sum_{s_1 = 0}^{s} \sum_{a_{1,\underline{\alpha}_{1}} + \ldots + a_{1,\underline{\alpha}_{s}} = s_1} \lambda(a_{1,\underline{\alpha}_{1}}, \ldots, a_{1,\underline{\alpha}_{s}}) \sum_{s_2 = 0}^{s - s_1} \sum_{a_{2,\underline{\alpha}_{1}} + \ldots + a_{2,\underline{\alpha}_{s}} = s_2} \lambda(a_{2,\underline{\alpha}_{1}}, \ldots, a_{2,\underline{\alpha}_{s}}) \ldots \nonumber \\
&\ldots& \sum_{s_n = s - s_1 - s_2 - \ldots - s_{n-1}} \sum_{a_{n,\underline{\alpha}_{1}} + \ldots + a_{n,\underline{\alpha}_{s}} = s_n} \lambda(a_{n,\underline{\alpha}_{1}}, \ldots, a_{n,\underline{\alpha}_{s}}) \nonumber \\
&=& \biggl( \sum_{s_1 = 0}^{\infty} y^{s_1} \sum_{\sum_{j = 1}^s a_{1,\underline{\alpha}_j} = s_1} \lambda(a_{1,\underline{\alpha}_{1}}, \ldots, a_{1,\underline{\alpha}_{s}}) \biggl) \ldots \biggl( \sum_{s_n = 0}^{\infty} y^{s_n} \sum_{\sum_{j = 1}^s a_{n,\underline{\alpha}_j} = s_n} \lambda(a_{n,\underline{\alpha}_{1}}, \ldots, a_{n,\underline{\alpha}_{s}}) \biggl) \nonumber \\
&=& \biggl( \sum_{s = 0}^{\infty} y^s \sum_{\sum_{j = 1}^s a_{1,\underline{\alpha}_j} = s_1} \lambda(a_{1,\underline{\alpha}_{1}}, \ldots, a_{1,\underline{\alpha}_{s}}) \biggl)^n
\end{eqnarray}
\end{widetext}

We observe that, when the expectations contain only one factor, for $s_1=1$,
we have \\
$E[h_{1,\underline{\alpha}_{1}}\!\!\!\!   ^{      a_{1,\underline{\alpha}_{1}} }
h_{1,\underline{\alpha}_{2}}\!\!\!\!   ^{      a_{1,\underline{\alpha}_{2}} }...]=E[h_{1,\underline{\alpha}_{j}}]=\lambda(a_{1,\underline{a}_j})=0$ for any $j$. 
As a consequence, the Taylor expansion of $f(y)$ in 
$y$ can be written as 
$f(y)=c_0+c_2y^2+c_3y^3+..$.\\ 
Eventually, it is the absence of $c_1y$ in this expansion 
that allows us to explain the properties of the polynomials in equation (\ref{esempio})
and finally the large sample approximation (\ref{5}).
To accomplish this last step we explain how the polynomials 
 $P_m(n)$ are related to the coefficients $c_i$.
Let us consider  the contribution  to $P_m(n)$ that originates 
from the product of $k$ factors $c_{i_1}y^{i_1},..,c_{i_k}y^{i_k}$
with the constraints  $  i_l \geq 1 $ and $i_1+..+i_k=m$. 
The number of ways these products can be formed 
for an assigned set of  $i_1,..,i_k$
is proportional to 
 $ \binom{n}{k}\simeq n^{k}+lower\,\,powers\,\, in \,\,n$.
Moreover, the contributions to $P_m(n)$ are formed by an integer number 
of factors less than or equal to $\frac{m}{2}$ because $c_1=0$. This 
property limits the highest power in $n$ contained in $P_m(n)$ with 
the largest integer number smaller than or equal to $\frac{m}{2}$.
Equation (\ref{esempio}) is then proved as is the large sample approximation (\ref{5}).\\

Each $asymptotic$ order $m$ of the $r^{th}$ component of the observed error 
is denoted by $( \widehat{\vartheta}-\vartheta )^r[m]$ where the index   
is given by  $m=\frac{s+1}{2}$ and $s$ is a natural number.
For example, the $asymptotic$ order for $m=1$
is given by 
\begin{eqnarray}
( \widehat{\vartheta }-\vartheta )^r [1]=\frac{1}{2}\upsilon^{rst}l_{s}l_{t}+i^{rs}i^{tu}H_{st}l_{u}  
\end{eqnarray}
where we have adopted the lifted indexes notation $\upsilon ^{a...z}=i^{a \alpha}...i^ {z \delta} \upsilon_{\alpha ..  \delta}$.
The $asymptotic$ orders of the bias are then given as the expectation 
of the $asymptotic$ orders of the observed error
\begin{equation}
\tilde{b}[ { \widehat{{\vartheta} }}^r , m ]=E[( \widehat{\vartheta }-\vartheta )^r[m]].
\end{equation} 
The $asymptotic$ order ${\tilde{b}}[ { \widehat{{\vartheta} }}^r , m ]$ contains different powers of $\frac{1}{n}$
as we discuss in this paragraph.\\
It follows from equations (\ref{2}) to (\ref{5})
that ${\tilde{b}}[ { \widehat{{\vartheta} }}^r , m ]$
is the sum of terms having the structure \\
${\underbrace{i^{(.)}\,\,..\,\,i^{(.)}}_{a}}\,\,{\underbrace{\upsilon_{(.)}\,\,..\,\,\upsilon_{(.)}}_{b}}\,\,E[{\underbrace{H_{(.)}\,\,..\,\,H_{(.)}}_{c}}]$
where $a$,$b$, and $c$ are the numbers of factors in the three products
 satisfying $a-b-\frac{c}{2}=m$.
Different powers of  $\frac{1}{n}$ can be generated because $ E[{\underbrace{H_{(.)}\,\,..\,\,H_{(.)}}_{c}}]$ can contain all the integer powers of $n$ equal to or less than $ n^{int(\frac{c}{2})}$.
However from the fact that no power higher than $ n^{int(\frac{c}{2})}$ can be generated  follows that an asymptotic order $m$ will never generate 
powers in the sample size $\frac{1}{n^p}$ with $p$ smaller than $m$.
It still remains to prove which is the largest power in $\frac{1}{n}$
contained in the asymptotic order $m$. We show that in the following.
Since the largest range of powers of $\frac{1}{n}$ is allowed when $b=0$,
we study these terms to evaluate the range of powers of $\frac{1}{n}$
contained in ${\tilde{b}}[ { \widehat{{\vartheta} }}^r , m ]$.
The structure of equation (\ref{2}) implies that its
 iteration with respect to the observed error components generates an equal number of 
$j^{ab}$ and $H_{(.)}$ (we recall that $l_{(.)}=H_{(.)}+\upsilon_{(.)}$).
Similarly, the number of $i^{(.)}$ generated from the expansion of $j^{ab}$  
given in equation (\ref{barnone}) is equal to  
the number of $H_{ab}$ plus 1.
These two observations imply that the terms where $b=0$ also verify $a=c=2m$. 
As a consequence, the powers of  $\frac{1}{n}$
that ${\tilde{b}}[ { \widehat{{\vartheta} }}^r , m ]$ contains can range from $the\,\, smallest\,\, integer\,\, number\,\, larger\,\, than\,\, or\,\, equal\,\, to\,\, m$, up to $2m-1$.\\
The analysis above implies that in order to compute
the contributions to  $\frac{1}{n}$ and to  $\frac{1}{n^2}$ in  the bias expansion 
that we denote with 
$b[ \widehat{\vartheta }^r, 1]$ and $b[ \widehat{\vartheta }^r, 2]$,  
it is necessary to obtain the first three $asymptotic$ orders of the bias,
${\tilde{b}}[  \widehat{\vartheta }^r ,1]$, 
${\tilde{b}}[  \widehat{\vartheta }^r ,\frac{3}{2}]$, and 
${\tilde{b}}[  \widehat{\vartheta }^r ,2]$.
In the explicit expressions below, 
to condense the notation, we introduce the
 quantities $I^{\alpha\beta,..,\gamma\delta}= i^{\alpha\beta}.. i^{\gamma\delta}$ , so that
\begin{widetext}
\begin{eqnarray}
{\tilde{b}}[  \widehat{\vartheta }^r , 1 ]&=&\label{11}I^{ri,kl}\upsilon _{ik,l}+\frac{1}{2}I^{rj,lp}\upsilon _{jlp}\\
{\tilde{b}}[  \widehat{\vartheta }^r ,\frac{3}{2}]&=&\frac{1}{6}(I^{r\alpha ,s\beta ,t\gamma ,u\delta }\upsilon _{\alpha\beta\gamma\delta}+\label{12}3I^{r\alpha ,s\beta ,\gamma\delta,tg,uv }\upsilon _{\alpha\beta\gamma}\upsilon _{\delta,g,v}
 \upsilon _{s,t,u}+ I^{r\alpha ,s\beta ,u\gamma ,tv }\upsilon _{\alpha\beta\gamma}\upsilon _{uv,s,t}+\\
&+&\frac{1}{2}I^{rs,t\alpha ,u\beta ,v\gamma}\upsilon _{\alpha\beta\gamma}\upsilon _{st,u,v} +
+\frac{1}{2}I^{rs,tu,vw}\upsilon _{stv,u,w} +I^{rs,tu,vw}\upsilon _{st,vu,w} +2I^{rs,tw}\upsilon _{st,w} \nonumber\\
&+&I^{r\alpha,\beta\gamma}\upsilon _{\alpha\beta\gamma}\nonumber\\
{\tilde{b}}[  \widehat{\vartheta }^r ,2]&=&\label{13}( I^{ra,bq,cd,tp}+  \frac{1}{2}I^{rd,ta,bq,cp}) 
 \upsilon_{\{abc\}\{dt\}\{q\}\{p\}}+\frac{1}{2}( I^{ra,bs}\upsilon ^{cpq}
 +I^{bp,cq}\upsilon ^{rsa}  ) \upsilon_{\{abc\}
 \{s\}\{p\}\{q\}} \nonumber \\
&+& (\frac{1}{2}I^{ap,tq}\upsilon ^{rbg}+I^{ra,tq}\upsilon ^{bpg}+
 \frac{1}{2}I^{ra,bg}\upsilon ^{tpq}+I^{bg,tq}\upsilon ^{rpa})\upsilon_{\{ab\}\{gt\}\{p\}\{q\}}+\frac{1}{2}[
 \upsilon ^{rsz}\upsilon ^{pqt} \nonumber\\
&+&\upsilon ^{rzqs}i^{tp}+\frac{1}{3}\upsilon ^{pzqt}i^{rs}+(I^{dz,eq,pt}\upsilon ^{rcs} 
 +I^{rp,dz,et}\upsilon ^{sqc}+2I^{dq,es,pt}\upsilon ^{rzc})\upsilon_{cde}]
\upsilon_{\{sp\}\{z\}\{q\}\{t\}}  \nonumber \\
&+&I^{ra,st,bc,de}\upsilon_{\{ab\}\{cd\}\{et\}\{s\}}+\frac{1}{6}I^{rj,kl,mp,qz}\upsilon_{\{jlpz\}\{k\}\{m\}\{q\}}+[\frac{1}{24}\upsilon^{r \alpha \beta \gamma \delta}+\nonumber\\
&+& \frac{1}{6}\upsilon^{r \alpha a} I^{\beta b,g\delta,\gamma d} \upsilon _{abgd}+\frac{1}{4}\upsilon^{r a \gamma \delta} I^{\beta c,,b \alpha} \upsilon_{a b c}+
  \frac{1}{8} \upsilon^{r v a} I^{w \alpha,z \beta, b \gamma, g \delta}       
 \upsilon_{v w z}\upsilon_{abg}+\nonumber\\
&+&\frac{1}{2}\upsilon^{r \alpha v} i^{w\beta}       
 \upsilon_{v w z}\upsilon^{z \gamma \delta}]\upsilon_{\{\alpha\}\{\beta\}\{\gamma\}\{\delta\}}. \label{BiasTwoRough}
\end{eqnarray}
\end{widetext}
The first order of the bias $b[ \widehat{\vartheta }^r, 1]$ can than be obtained by substituting 
the explicit expressions for the tensors in ${\tilde{b}}[  \widehat{\vartheta }^r ,1]$. 
The second order $b[ \widehat{\vartheta }^r, 2]$
 takes contributions
from ${\tilde{b}}[  \widehat{\vartheta }^r ,\frac{3}{2}]$
and ${\tilde{b}}[  \widehat{\vartheta }^r ,2]$. However, while ${\tilde{b}}[  \widehat{\vartheta 
}^r ,\frac{3}{2}]$ generates only $n^{-2}$ contributions, ${\tilde{b}}[  
\widehat{\vartheta }^r ,2]$ generates $n^{-2}$ and $n^{-3}$ contributions. 
Consequently, to collect all and only the contributions to $n^{-2}$,
it is necessary to extract the $n^{-2}$ component from ${\tilde{b}}[  \widehat{\vartheta }^r ,2]$ 
and add it to ${\tilde{b}}[  \widehat{\vartheta }^r ,\frac{3}{2}]$.
The extraction can be done by introducing into ${\tilde{b}}[  \widehat{\vartheta }^r ,2]$ only the highest 
powers of $n$ of the tensors. 

The first two orders of the bias for the MLE of the $r$ component of the 
parameter vector $\underline{\vartheta}$ then become (\ref{BiasOne}) and (\ref{BiasTwo})

Starting from (\ref{BiasTwo}) the form for the second order covariance matrix can be obtained using the following theorem:\\
Theorem: {\it
 Let $g(.)$ be an invertible and differentiable function 
and let $\underline{\xi}=g(\underline{\vartheta})$ be a set of parameters 
dependent on $\underline{\vartheta}$.
Let $b(\widehat{\vartheta }_r)$ and  $b(\widehat{\xi }_r)$ be
the biases relative to the estimation of the
components of $\underline{\vartheta}$ and $\underline{\xi}$
which can be computed as power series of $\frac{1}{n}$, as explained in Section 2.  
 The terms of the expansion in powers of $\frac{1}{n}$
for $b(\widehat{\xi }_r)$
can be obtained from the terms of the expansion for $b(\widehat{\vartheta }_r)$
 by replacing the derivatives with respect to the
components of $\underline{\vartheta}$, with derivatives with respect to the components of
$\underline{\xi}$. 
The explicit dependence on $\underline{\xi}$ can be removed at the end
of a derivation by means of the Jacobian matrix J defined by 
$\frac{\partial }{\partial \xi_m}=\frac{\partial \vartheta_s}{\partial \xi_m}\frac{\partial }{\partial \vartheta_s} = J_{ms}\frac{\partial }{\partial \vartheta_s}$ .
 }	\\

To prove the chain-rule theorem given in section \ref{Expansions} , we analyze
how the derivation of the expansion in powers of $\frac{1}{n}$
for $b(\vartheta_r)\,\, r=1,..,D$, which is described there,
changes if we are interested in the expansion for $b(\xi_r)\,\, r=1,..,D$ where  
$ \underline{\xi}$ is a vector of parameters in an invertible relationship 
with ${\underline{\vartheta}}$: ${\underline{\xi}}=f(\underline{\vartheta}); \,\,\,\,\underline{\vartheta}=f^{-1}({\underline{\xi}})$.
The starting point for the derivation of the expansion of $b(\xi_r)\,\, r=1,..,D$ 
is the stationarity condition 
\begin{eqnarray}
  l_r(\underline{x},\underline{\widehat{\xi}}) =\frac{\partial l(\underline{x},f^{-1}(\underline{\xi}))}{\partial\xi_r }{{\mid}_{\underline{\xi}=\underline{\widehat{\xi}}}}=0 \,\,\,\,\,\,\, r=1,..,D.
\end{eqnarray}
that can be obtained from equation (\ref{stazionarieta}) by replacing 
only the derivatives $\frac{\partial }{\partial \vartheta_r}\,\,for \, \, r=1,..,D$
with  $\frac{\partial }{\partial \xi_r}\,\,for \,\, r=1,..,D$, because 
$\underline{\vartheta}=\underline{\widehat{\vartheta}}$
implies $\underline{\xi}=f(\underline{\widehat{\vartheta}})=
\underline{\widehat{\xi}}$.
The subsequent steps can then be obtained from
equations  (\ref{cippo}) up to (\ref{BiasTwo}) 
by replacing the derivatives in the same way.
Since 
$\frac{\partial}{\partial \xi_m}=\frac{\partial\vartheta_s}{\partial \xi_m}\frac{\partial}{\partial\vartheta_s}=J_{ms}\frac{\partial}{\partial\vartheta_s}$,
 where the components of the Jacobian matrix $J_{ms}$ behave like constants in the expectations, the substitution 
of the derivatives can also be done only in the final result (for example, in the 
orders given in equations (\ref{BiasTwoRough}) and  (\ref{BiasTwo})).
The expectations contained in the expansion of $b(\vartheta_r)$ in powers of $\frac{1}{n}$ can also be computed before the substitution 
of the derivatives if the likelihood function dependence on the parameters 
is expressed in terms of 
auxiliary functions. Examples of auxiliary functions are $g_r(\underline{\vartheta})= \vartheta_r$
for a general parametric dependence and $g_1(\vartheta)=\mu(\vartheta)$
$g_2(\vartheta)=\sigma(\vartheta)$ for a scalar Gaussian distribution. 
By means of these 
auxiliary functions, the derivatives  $\frac{\partial}{\partial \vartheta_m}$ and 
$\frac{\partial}{\partial \xi_m}$ become 
$\frac{\partial g_p }{\partial \vartheta_m} \frac{\partial}{\partial g_p}$
and 
$ \frac{\partial g_p}{\partial \xi_m} \frac{\partial}{\partial g_p}$.
As a consequence the orders of the expansion for $b(\xi_r)$
can be found from the orders for the expansion of $b(\vartheta_r)$
implementing, in the result of the expectations, the substitutions
\begin{equation}
\frac{{\partial}^{i_1+i_2+...+i_p} g_m(\underline{\vartheta})}
{({\partial\vartheta}_1)^{i_1}..({\partial\vartheta}_D)^{i_D}}\rightarrow
\frac{{\partial}^{i_1+i_2+...+i_D} g_m (f^{-1}(\xi))}{({\partial\xi}_1)^{i_1}..({\partial\xi}_D)^{i_p}.}
\end{equation}
The converse of the chain-rule, in which higher moments of the observed error are used to compute lower moments, 
is not possible.
We can observe, for example, that the expansion 
of a general moment of an MLE does not always contain all the powers of $\frac{1}{n}$.
The lowest order present in the expansion of the m order moment
is given by the largest integer number smaller 
than or equal to $\frac{m}{2}$ (we use the notation $int(\frac{m}{2})$):
\begin{eqnarray*} 
E[(\widehat{\vartheta }-\vartheta)^{i_{1}}
...
(\widehat{\vartheta }-\vartheta)^{i_{m}}]= \frac{1}{n^{int(\frac{m}{2})}}+higher\,powers\, o\!f\, \frac{1}{n}.
\end{eqnarray*}
The consequence of this observation is that only the bias and the error correlation matrix 
may contain first order terms.
For this reason an inverse chain-rule would have 
to generate non-zero orders of lower moments expansions
from the corresponding orders of the higher moments expansions that are zero
for powers of $\frac{1}{n}$ lower than $\frac{1}{n^{int(\frac{m}{2})}}$.  

Let us consider how the chain-rule  makes it possible 
 to compute the expansion of the 
error correlation matrix $\Xi(\widehat{\vartheta })$ defined by 
$\Xi(\underline{\widehat{\vartheta }})=E[(\widehat{\underline{\vartheta} }-\underline{\vartheta})(\widehat{\underline{\vartheta} }-\underline{\vartheta})^T]$ and the covariance matrix.
Using the invariance property of the MLE,
\begin{eqnarray}
\Xi(\underline{\widehat{\vartheta }})=b(\widehat{{\underline{\vartheta}{\underline{\vartheta}}^T}})-\underline{\vartheta} \label{74}(b({\underline{\widehat{\vartheta }} }))^T-b({\underline{\widehat{\vartheta }} })
{\underline{\vartheta} }^T
\end{eqnarray}
where $(.)^T$ is the transpose of a vector and 
 $b(\widehat{{\underline{\vartheta}{\underline{\vartheta}}^T}})$ is a matrix whose components 
are the bias of the product of two components of $\underline{\vartheta}$.
Once $\Xi(\underline{\widehat{\vartheta }})$ and the bias are known,  
then the covariance matrix $C(\widehat{\vartheta })$  can also be computed 
by means of
\begin{eqnarray}
 C(\widehat{\vartheta })=\Xi(\widehat{\vartheta })\label{conversion}-b(\underline{\widehat{\vartheta }})b(\underline{\widehat{\vartheta }})^{T}.
\end{eqnarray}
To compute the right hand side of equation (\ref{74})
we express it in terms of the components, obtaining 
\begin{eqnarray}
\Xi_{ij}(\underline{\widehat{\vartheta }})=b(\widehat{{\vartheta_i}{\vartheta}_j})-\label{21}\vartheta_i b({\widehat{\vartheta } }_j)
-b({\widehat{\vartheta }_i }){\vartheta }_j.
\end{eqnarray} 
It is important to realize that knowledge of 
$b({{\widehat{\vartheta } }}_r)$ is sufficient because 
the expansion of $b(\widehat{{\vartheta_i}{\vartheta}_j})$ can be derived 
from it by means of the theorem given above.
In fact, if  we choose
\begin{eqnarray}  
\underline{\xi}=\label{variab} \{ \vartheta_1,..,\vartheta_{i-1},\vartheta_{i}\vartheta_{j},\vartheta_{i+1},..,\vartheta_{D}\} 
\end{eqnarray}
as a new set of parameters,  
$b(\widehat{{\vartheta_i}{\vartheta}_j})$ becomes 
 $b(\widehat{{\xi_i}})$. However, is necessary to insure that  
the relationship between $\underline{\vartheta}$ and $\underline{\xi}$ is invertible. 
This condition holds 
if both $\vartheta_{i}\neq0 $ where   $ i=1,..,D$ and
the sign of the components of $\underline{\vartheta}$ is known,
for example, by means of constraints on the data model.\\
In a scalar estimation scenario, equation (\ref{74}) becomes simply 
\begin{eqnarray}
\Xi(\widehat{\vartheta })=\label{simplechain}E[(\widehat{\vartheta }-\vartheta)^2]=b(\label{61}\widehat{{\vartheta }^2})-2\vartheta b(\widehat{{\vartheta }})
\end{eqnarray}
and the variance $C(\widehat{\vartheta })=\Xi(\widehat{\vartheta })-
b(\underline{\widehat{\vartheta }})^2$.
In this case $b(\widehat{{\vartheta }^2})$ can be derived from $ b(\widehat{{\vartheta }})$
if we choose $\xi=\vartheta^2$ as the new parameter and the Jacobian matrix becomes  
 $\frac{1}{2\vartheta}$  because  $\frac{\partial}{\partial \xi}=
\frac{\partial \vartheta }{\partial \xi} \frac{\partial}{\partial \vartheta}
=\frac{1}{2\vartheta}\frac{\partial}{\partial \vartheta}$.\\ 
A useful simplification of the algebra of 
the chain-rule  in the derivation of second order moments 
is described in the following two paragraphs.
The chain-rule  and the subsequent conversion of the derivatives
require the substitutions
\begin{widetext}
\begin{eqnarray}
\frac{ \partial^m [.] }{ \partial \vartheta_{i_1} ..\partial \vartheta_{i_m}  } \rightarrow
\frac{ \partial^m [.] }{ \partial \xi_{i_1}.. \partial \xi_{i_m}  }  \rightarrow
J_{i_1 j_1}(\underline{\vartheta}) \frac{ \partial }{ \partial \label{convder} \vartheta_{j_1} }
J_{i_2 j_2}(\underline{\vartheta}) \frac{ \partial }{ \partial \vartheta_{j_2} }..
J_{i_m j_m}(\underline{\vartheta}) \frac{ \partial }{ \partial \vartheta_{j_m} }[.]
\end{eqnarray}
\end{widetext}
From the right hand side of (\ref{convder}) it is clear that 
the derivatives will also be applied to the Jacobian matrix,
thereby generating $m!$ terms for every derivative of order  $m$. 
However, it can be observed that the terms contributing to the final result 
are only a small subset.
In fact, among all the  terms generated from the conversion of the derivatives 
in the bias expansion, 
only those where a first derivative of the Jacobian matrix appears must be considered.
For example, we can consider the term 
$\frac{1}{2}I^{ra,sb,vc}\upsilon_{abc}\upsilon _{v,w}I^{wd,te,uf}\upsilon_{def}
\upsilon _{s,t,u}$ that comes from equation  (\ref{BiasTwo}), which must be used 
to derive the second order $\Xi(\widehat{\vartheta })$. In this case,
we need to consider only 
the 3 terms in which one of the derivatives represented 
by $a, b, c$ operates on a Jacobian matrix and 
$d,e, f$ operate on the likelihood function, plus
the 3 terms where the role of $a,b,c$ and $d,e,f$ are inverted.
In general, the total number of relevant terms  
among those generated in every derivative is equal to or 
less than the  order of the derivative $m$.\\ 
The detailed analysis of 
equation (\ref{21}) reveals that 
the terms generated  
in $bias(\widehat{\vartheta_i{\vartheta}_j})$      
can be divided into three groups: (a)
the terms where no derivative of the Jacobian matrix appears 
are equal to ${((J^{-1})^T)_{is}}bias(\widehat{\vartheta }_s)$
(we show in Example 1 that  ${((J^{-1})^T)_{is}}bias(\widehat{\vartheta }_s)$  cancels 
with      $-\vartheta_i b({\widehat{\vartheta } }_j)
-b({\widehat{\vartheta }_i }){\vartheta }_j$ after its introduction in equation (\ref{21}));
(b) the terms where only one derivative of the Jacobian matrix 
appears give the error correlation matrix;
and (c) the terms that contain higher derivatives of the Jacobian matrix
 or more than one first derivative of the Jacobian matrix in the same term, summed together, give zero.\\
\\
To clarify the use of the chain-rule  and the algebraic issues discussed above, we present 
two examples. 
In Example 1 we use the first order term of the bias in a general vector estimate
 to derive the first order term of $\Xi(\widehat{\vartheta })$.
It is useful to recall that the expansion of $\Xi(\widehat{\vartheta })$ and the expansion of $C(\widehat{\vartheta })$
can be different only from the second order on, so this example also describes a derivation 
of the first order of the covariance matrix $C(\widehat{\vartheta })$.
Following the same approach, the second order term of the error correlation matrix 
expansion can be derived from equation (\ref{BiasTwo}) and the second order 
covariance matrix can also be derived if we also use (\ref{BiasOne}) and (\ref{conversion}).\\
In Example 2 we illustrate the way the chain-rule  can still be used, if the available expression  
 of the bias expansion is explicitly computed for a particular data model. 
In particular, we derive  
the second order  mean square error and the second order variance from the second 
order bias in two scalar Gaussian models.
In Example 2 we also illustrate the simplification introduced above  for the 
algebra involved in the conversion 
of the derivatives.\\ 
\\
Example 1\\
Using the Bartlett identities (\cite{Bartlett1953},\cite{Bartlett1953-}), we rewrite the first order bias, 
given by equation ($\ref{11}$), as 
\begin{eqnarray}
b[\widehat{\vartheta }_r, 1]=-\frac{1}{2}i^{rj}i^{lp}\label{trichecone}(\upsilon _{j,l,p}+\upsilon _{j,lp})
\end{eqnarray}
From equation (\ref{trichecone}), $b(\widehat{\vartheta_i{\vartheta}_j})=b(\xi_i)$
can be computed by means of the chain-rule  
by replacing the derivatives with respect to the components of 
$\underline{\vartheta}$ with derivatives with respect to the components of  
$\underline{\xi}$, where $\underline{\xi}$ is given in equation (\ref{variab}) and using the corresponding Jacobian matrix. The tensors in equation (\ref{trichecone}) become
\begin{eqnarray*}
i_{lp}(\xi)&=&E[ \frac{\partial^2 l}{ \partial \xi_l \partial \xi_p} ]=J_{lr}i_{rs}(\vartheta)(J^T)_{sp}=\\ &=& J_{lr}E[ \frac{\partial^2 l}{ \partial \vartheta_i \partial \vartheta_j}]\label{tricheco2}(J^T)_{sp}\\	
{i^{lp}}(\xi)&=&{((J^{-1})^T)_{lr}}{i^{rs}}(\vartheta)(J^{-1})_{sp}\\
\upsilon _{j,l,p}\label{tricheco3}(\xi)&=&J_{j\alpha}J_{l\beta}J_{p\gamma}\upsilon _{\alpha,\beta,\gamma}(\vartheta)\\
\upsilon _{j,lp}(\xi)&=&\label{tricheco4}J_{j\alpha}J_{l\beta}(J_{p\gamma}\upsilon _{\alpha,\beta,\gamma}(\vartheta)+
\frac{\partial J_{p \gamma}}{\partial \beta} \upsilon _{\alpha,\gamma}(\vartheta))
\end{eqnarray*}
where we have specified in the bracket beside the tensors 
the dependence on the parameter sets.
Inserting these expressions in equations (\ref{trichecone}) and
observing that equation (\ref{21}) can be expressed 
in the form 
\begin{eqnarray}
 \Xi(\underline{\widehat{\vartheta }})_{ij}=b(\widehat{\vartheta_i{\vartheta}_j})-\label{23}{((J^{-1})^T)_{is}}b(\widehat{\vartheta }_s)
\end{eqnarray}
the first order term of the error correlation matrix
can then be obtained as
\begin{eqnarray}
\Xi_{ij}[\underline{\widehat{\vartheta }},1]=\frac{1}{2} i^{\alpha \beta}\frac{\partial 
(J^{-1})_{\alpha i}}{\partial \vartheta_{\beta}}=\frac{1}{2} i^{\alpha \beta}\gamma_{\alpha \beta}=i^{ij}
\end{eqnarray}
where we have introduced the tensor
\[ \gamma_{\alpha \beta}= \left \{ \begin{array}{ll}
 1  & \mbox{if $\alpha=i\,\, \beta=j$ or $\alpha=j\,\, \beta=i$}\\
0  & \mbox{otherwise.$\,\,\,\,\,\,\,\,\,\,\,\,\,\,\,\,\,\,\,\,\,\,\,\,\,\,\,\,\,\,\,\,\,\,\,\,\,\,\,\,\,\,\,\,\,\,\,\,\,\,\,\,\,\,\,\,\,\,\,\,\,\,\,\,$}
                \end{array}
        \right. \]  \\
Example 2\\
In this example 
we determine the second order variance and mean square error from the second order bias
for two cases of parameter dependence for the scalar Gaussian density
\begin{eqnarray}
p(x,\vartheta)=\label{gausca}\frac{1}{(2 \pi c)^{n/2}}\exp\left(-\frac{1}{2}\sum_{i=1}^{n}\frac{(x_i-\mu)^2}{c}\right) 
\end{eqnarray}
as a direct application of equation (\ref{simplechain}).\\
In the case where c does not depend on the parameters ($ \frac{\partial c} {\partial\vartheta} = 0 $)
the second order bias can be derived using the 
scalar version of equation   (\ref{BiasTwo}) and of the tensors $\upsilon$ defined in equation (\ref{UpsilonDef}).
For this parameter dependence, 
the $asymptotic$ order for $m=\frac{3}{2}$  (equation (\ref{12}))
is zero, and the second order bias can  be directly obtained also from equation (\ref{13}) 
 The result is $b[\frac{1}{n^2}]=\frac{c^2}{(\dot{\mu})^8}[\frac{5}{4}\dddot{\mu}\ddot{\mu}(\dot{\mu})^2-\frac{1}{8}\ddddot{\mu}(\dot{\mu})^3-\frac{15}{8}\dot{\mu}(\ddot{\mu})^3]$.
Applying the chain-rule, the second order mean square error for the estimation of 
$\vartheta$ becomes $\Xi(\vartheta)[\frac{1}{n^2}]=\frac{15{\ddot{\mu}}^2}{ 4{\dot{\mu}}^6 }-\frac{\dddot{\mu}}{{\dot{\mu}}^5 }$, where the full conversions of the derivatives are given by  
$\dot{\mu}\!\rightarrow\!\frac{\partial \mu}{\partial {\vartheta}^2}\!=\!\frac{\dot{\mu}}{ 2\vartheta }\,\,;\,\,\ddot{\mu}\!\rightarrow\! \frac{{\partial}^2 \mu}{(\partial {\vartheta}^2)^2}\!=\!\frac{1}{ 2\vartheta }(\frac{\ddot{\mu}}{ 2\vartheta }-\frac{\dot{\mu}}{ 2{\vartheta}^2 })\,\,;\,\,\dddot{\mu}\!\rightarrow\!\frac{{\partial}^3 \mu}{(\partial {\vartheta}^2)^3}\!=\!\frac{1}{ 2\vartheta }(\frac{\dddot{\mu}}{ 4{\vartheta}^2 }-\frac{3\ddot{\mu}}{ 4{\vartheta}^3 }+\frac{3\dot{\mu}}{ 4{\vartheta}^4 })\,\,;\,\, 
\ddddot{\mu}\!\rightarrow\! \frac{{\partial}^4 \mu}{(\partial {\vartheta}^2)^4}\!=\!\frac{\ddddot{\mu}}{ 16{\vartheta}^4 }-\frac{3\dddot{\mu}}{ 8{\vartheta}^5 }+\frac{15\ddot{\mu}}{ 16{\vartheta}^6 } -\frac{15\dot{\mu}}{ 16{\vartheta}^7 }$. By means of these conversions and the expression of the second 
order bias, it can be observed that among the 18 terms that are 
in principle generated by the chain-rule, only 6 contribute 
to the second order mean square error. \\
\\ 
In the following we show that the expansion in the inverse of the sample size is equivalent to an expansion in $1/SNR$.
The derivation of the asymptotic orders in $\frac{1}{n}$ 
would be the same for an expansion in any quantity if 
(\ref{UpsilonDef}) and (\ref{5}) can be derived for a certain quantity 
$\gamma$ instead of the sample size n.
In this section we illustrate indeed that this is the case where
the signal to noise ratio for a set of scalar data distributed according
to a deterministic Gaussian PDF takes the role of the sample size.
The probability distribution and parameter dependent part of the likelihood function are given by:
\begin{eqnarray}
p(x,\vartheta) &=& \frac{1}{{(2\pi\sigma^2)^{n/2}}}e^{-\frac{\sum_{i = 1}^{n}(x_i - \mu(\vartheta))^2}{\sigma^2}} \\
l(\vartheta) &=& -\frac{\sum_{i = 1}^{n}(x_i - \mu(\vartheta))^2}{\sigma^2}
\end{eqnarray}
We also define the signal to noise ratio for this example as $\gamma = {\mu^2}/{\sigma^2}$ following standard practice for scalar deterministic 
Gaussian data. We can obtain:
\begin{widetext}
 \begin{eqnarray}
H_\alpha(\vartheta) &=& 2\mu_\alpha(\vartheta)\frac{\sum_{i = 1}^{n}(x_i - \mu(\vartheta))}{\sigma^2} \\
l_\alpha(\vartheta) &=& -n\sum_p c(p) \frac{\mu^p\mu^{\alpha - p}}{\sigma^2} + 2\mu_\alpha(\vartheta) \frac{\sum_{i = 1}^{n}(x_i - \mu(\vartheta))}{\sigma^2} \\
\upsilon_\alpha &=& <l_\alpha(\vartheta)> = \label{gaussian_expect_la} n\frac{\mu^2}{\sigma^2}[-\sum_p c(p) (\frac{\mu^p}{\mu}) (\frac{\mu^{\alpha - p}}{\mu})]
\end{eqnarray}
\end{widetext}

\noindent where $\alpha$ or p denote the order of the derivative or an arbitrary set of derivatives.

From (\ref{gaussian_expect_la}) it becomes obvious that $\upsilon_\alpha$ is proportional to not 
only the sample size, $n$, but also to the $\gamma$. The term inside the square brackets is simply a 
sum over normalized derivatives of the mean and contains information about the shape of the signal. 
In the above equations we use:

\begin{eqnarray}
E[(x - \mu(\vartheta))^{2n}] = \sigma^{2n}\frac{(2n)!}{2^n n!}
\end{eqnarray}

The next step is to determine the dependence of $E[H_{\alpha_1} \ldots H_{\alpha_p}]$ on $\gamma$. This 
is shown below:

\begin{eqnarray}
E[H_{\alpha_1} \ldots H_{\alpha_p}] &=& 2^p \mu_{\alpha_1} \ldots \mu_{\alpha_p} \frac{1}{\sigma^{2p}} E[\sum_{i = 1}^{n}(x_i - \mu(\vartheta))^p] \nonumber \\
&=& 2^p (\frac{\mu_{\alpha_1}}{\mu}) \ldots (\frac{\mu_{\alpha_p}}{\mu}) \frac{\mu^p}{\sigma^{2p}} \sigma^p \frac{(2\frac{p}{2})!}{2^{\frac{p}{2}}(\frac{p}{2})!}n^{\frac{p}{2}} \nonumber \\
&=& (\frac{\mu^2}{\sigma^2})^{\frac{p}{2}} \ldots (\frac{\mu_{\alpha_1}}{\mu}) \ldots (\frac{\mu_{\alpha_p}}{\mu})n^{\frac{p}{2}} \nonumber \end{eqnarray}
\begin{eqnarray}
\Rightarrow E[H_{\alpha_1} \ldots H_{\alpha_p}] &\propto& (n\gamma)^{\frac{p}{2}}
\end{eqnarray}

It is therefore found that $E[H_{\alpha_1} \ldots H_{\alpha_p}]$ is proportional to $(\gamma)^{\mu/2}$ 
and we have proved the analogy between sample size and $\gamma$.

Note that for deterministic Gaussian data, the non-integer asymptotic orders of the bias are zero and 
the integer orders are equal  to $b_1, b_2$ etc. This is sufficient to prove that the orders in 
$1/n$ of the bias expansion are also orders in $1/\gamma$. This also holds for the variance expansion.
A similar, although longer proof can be written for the SNR definition provided in \ref{SNRtoA}.

\end{document}